\documentstyle[preprint,eqsecnum,aps,emlines2,bezier]{revtex}
 \tightenlines

\def\eps{\varepsilon}
\def\partt{\mbox{\boldmath $\partial$}}
\def\const{{\rm const\,}}
\def\Dm{\widetilde{\cal D}_{\mu}}
\def\D{{\cal D}}
\def\h{{\bf h}}
\def\gdot{\circle*{20}}

\def\dhline#1#2#3#4#5#6{  {\countdef\nnn=255
\dimendef\llx=0   \dimendef\lly=1
\dimendef\dx=2   \dimendef\dy=3
\llx=#1\unitlength  \lly=#2\unitlength
\dx=#4\unitlength   \dy=#5\unitlength  \nnn=#6
\divide\nnn by 2
\advance\dx by-\llx \advance\dy by-\lly
\div\nnn \div4 \lline \adv
\multiply\dx by2 \multiply\dy by2
\loop \adv \ifnum\nnn>1 \lline \adv \advance\nnn by-1
\repeat \div2 \lline }}

\def\div#1{ \divide\dx by#1  \divide\dy by#1 }
\def\adv{ \advance\llx by\dx \advance\lly by\dy }

\def\lline{ {  \divide\llx by\unitlength \divide\lly by\unitlength
\divide\dx by\unitlength \divide\dy by\unitlength
\advance\dx by\llx \advance\dy by\lly
\emline{\number\llx}{\number\lly}{}{\number\dx}{\number\dy}{}}}

\def\dA{
\unitlength=0.04ex
\special{em:linewidth 0.3pt}
\begin{picture}(180,140)
\emline{10}{10}{}{90}{130}{}
\emline{90}{130}{}{170}{10}{}
\dhline{20}{25}{}{160}{25}{10}
\emline{38}{39}{}{26}{47}{}
\emline{154}{47}{}{142}{39}{}
\put(90,130){\gdot}
\end{picture}}

\def\dS{
\unitlength=0.04ex
\special{em:linewidth 0.3pt}
\begin{picture}(180,40)
\emline{10}{70}{}{170}{70}{}
\emline{23}{63}{}{23}{77}{}
\emline{130}{63}{}{130}{77}{}
\dhline{36}{70}{}{45}{82}{2}
\dhline{45}{82}{}{57}{92}{2}
\dhline{57}{92}{}{70}{99}{2}
\dhline{70}{99}{}{85}{103}{2}
\dhline{85}{103}{}{101}{103}{2}
\dhline{101}{103}{}{116}{99}{2}
\dhline{116}{99}{}{129}{92}{2}
\dhline{129}{92}{}{141}{82}{2}
\dhline{141}{82}{}{150}{70}{2}
\end{picture}}

\def\areas{
\unitlength=1.00mm
\special{em:linewidth 0.4pt}
\linethickness{0.4pt}
\begin{picture}(150.00,150.00)
\put(45.00,150.00){\vector(0,-1){125.00}}
\put(5.00,110.00){\vector(1,0){135.00}}
\emline{90.00}{150.00}{5}{90.00}{88.00}{6}
\put(18.00,65.00){\framebox(10.00,6.00)[cc]{FPI}}
\put(55.00,65.00){\framebox(10.00,6.00)[cc]{FPI}}
\put(101.00,65.00){\framebox(11.00,6.00)[cc]{FPII}}
\put(18.00,131.00){\framebox(12.00,6.00)[cc]{FPIII}}
\put(61.00,131.00){\framebox(13.50,6.00)[cc]{FPIV$\!a$}}
\put(106.00,131.00){\framebox(13.50,6.00)[cc]{FPIV$b$}}
\put(45.00,22.00){\makebox(0,0)[cc]{$\eta$}}
\put(138.00,113.00){\makebox(0,0)[cc]{$\varepsilon$}}
\put(113.00,77.00){\circle*{1.60}}
\put(133.00,64.00){\makebox(0,0)[cc]{$\varepsilon=2\eta$}}
\put(96.00,107.00){\makebox(0,0)[cc]{$\varepsilon=2$}}
\put(120.00,85.00){\makebox(0,0)[cc]{$\varepsilon=2\eta=8/3$}}
\put(125.00,26.00){\makebox(0,0)[cc]{$\varepsilon=\eta$}}
\special{em:linewidth 1.4pt}
\linethickness{1.4pt}
\emline{45.00}{110.00}{1}{120.00}{30.00}{2}
\emline{45.00}{110.00}{3}{131.00}{68.00}{4}
\emline{45.00}{110.00}{9}{45.00}{150.00}{10}
\emline{45.00}{110.00}{7}{4.00}{110.00}{8}
\end{picture}
\centerline{Fig.I. Regions of stability for the fixed points in
the model (2.4).} }

\begin{document}
\draft

 \title{Anomalous scaling regimes of a passive scalar
 advected by the synthetic velocity field}

 \author{N. V. Antonov}

 \address{Department of Theoretical Physics, St~Petersburg University,
 Uljanovskaja 1, \\ St~Petersburg, Petrodvorez, 198904 Russia}


 \maketitle

 \begin{abstract}
The field theoretic renormalization group (RG) is applied to the
problem of a passive scalar advected by the Gaussian self-similar
velocity field with finite correlation time and in the presence
of an imposed linear mean gradient. The energy spectrum in the
inertial range has the form $E(k)\propto k^{1-\eps}$, and the
correlation time at the wavenumber $k$ scales as $k^{-2+\eta}$. It
is shown that, depending on the values of the exponents $\eps$
and $\eta$, the model in the inertial-convective range exhibits
various types of scaling regimes associated with the infrared stable
fixed points of the RG equations: diffusive-type regimes for which
the advection can be treated within ordinary perturbation theory,
and three nontrivial convection-type regimes for which the correlation
functions exhibit anomalous scaling behavior. Explicit asymptotic
expressions for the structure functions and other correlation
functions are obtained; they are represented by superpositions of
power laws with nonuniversal amplitudes and universal (independent of
the anisotropy) anomalous exponents, calculated to the first order in
$\eps$ and $\eta$ in any space dimension. These anomalous exponents
are determined by the critical dimensions of tensor composite operators
built of the scalar gradients, and exhibit a kind of hierarchy related
to the degree of anisotropy: the less is the rank, the less is the
dimension and, consequently, the more important is the contribution
to the inertial-range behaviour. The leading terms of the even (odd)
structure functions are given by the scalar (vector) operators. For
the first nontrivial regime the anomalous exponents are the same as in
the rapid-change version of the model; for the second they are the
same as in the model with time-independent (frozen) velocity field.
In these regimes, the anomalous exponents are universal in the sense
that they depend only on the exponents entering into the velocity
correlator. For the last regime the exponents are nonuniversal
(they can depend also on the amplitudes); however, the nonuniversality
can reveal itself only in the second order of the RG expansion.
A brief discussion of the passive advection in the non-Gaussian
velocity field governed by the nonlinear stochastic Navier-Stokes
equation is also given. St Petersburg University Preprint SPbU
IP-98-16; {\it chao-dyn/9808011}; accepted to Phys.~Rev.~E.
 \end{abstract}


 \section{Introduction} \label {sec:Int}

The investigation of intermittency and anomalous scaling in
fully developed turbulence remains one of the major theoretical
problems. Both the natural and numerical experiments suggest
that the deviation from the predictions of the classical
Kolmogorov--Obukhov theory is even more strongly pronounced for a
passively advected scalar field than for the velocity field itself;
see, e.g., \cite{An,Sree,synth,pass1,pass2,El} and literature
cited therein. At the same time,
the problem of passive advection appears to be easier tractable
theoretically: even simplified models describing the advection by
a ``synthetic'' velocity field with prescribed Gaussian statistics
reproduce many of the anomalous features of genuine turbulent
heat or mass transport observed in experiments, see
\cite{synth}--\cite{Eyink}. Therefore, the
problem of a passive scalar advection, being of practical
importance in itself, may also be viewed as a starting point
in studying anomalous scaling in the turbulence on the whole.

Recently, a great deal of attention has been drawn by a simple
model of the passive scalar advection by a self-similar Gaussian
white-in-time velocity field, the so-called ``rapid-change model,''
introduced by Kraichnan \cite{Kraich1}; see [8--30]
and references therein. For the first time, the anomalous exponents
have been calculated on the basis of a microscopic model and within
regular expansions in formal small parameters. Within the
``zero-mode approach'' to the rapid-change model, developed in
\cite{Falk1,Falk2,GK,BGK}, nontrivial anomalous
exponents are related to the zero modes (homogeneous solutions)
of the closed exact equations satisfied by the equal-time
correlations. In this sense, the model is ``exactly solvable.''
The anomalous exponents are universal, i.e., they depend only
on the space dimension and the exponent entering into
the velocity correlator.

Of course, the Gaussian character, isotropy, and time decorrelation
are strong departures from the statistical properties of genuine
turbulence. One step toward the construction of a more realistic
model of passive advection is the account of the finite
correlation time of the velocity field.

In \cite{ShS,ShS2}, a generalized phenomenological model was considered
in which the temporal correlation of the advecting field was set by
eddy turnover (see also an earlier work \cite{PDF}, where the
probability distribution function in an analogous model was studied).
It was argued that the anomalous
exponents may depend on more details of the velocity statistics,
than only the exponents. This idea has received some analytical
support in \cite{Falk3}, where the case of short but finite correlation
time was considered for the special case of a local turnover exponent.
The anomalous exponents were calculated within the perturbation theory
with respect to the small correlation time, with Kraichnan's
rapid-change model taken as the zeroth order approximation.
The exponents obtained in \cite{Falk3} appear to be nonuniversal,
through the dependence on the correlation time. The exact inequalities
obtained in \cite{Eyink} using the so-called refined similarity
relations also point up some significant differences between the zero
and finite correlation-time problems.

In the paper \cite{RG}, the field theoretic renormalization group (RG)
and operator product expansion (OPE) were applied to the model
\cite{Kraich1}. The feature specific to the theory of turbulence is the
existence in the corresponding field theoretical models of the
composite operators with {\it negative} scaling (``critical'') dimensions.
Such operators are termed ``dangerous,'' because their contributions to
the OPE  for the structure functions
and various pair correlators give rise the anomalous scaling, i.e.,
singular dependence on the IR scale with nonlinear anomalous exponents.
The latter are determined by the critical dimensions of these
operators.\footnote{For the rapid-change model, the relationship
between the anomalous
exponents and dimensions of composite operators was anticipated
in \cite{Falk2,BGK,Eyink} within certain phenomenological formulation
of the OPE, the so-called ``additive fusion rules,'' typical
to the models with multifractal behavior; see also \cite{DL,Ey}.
The RG analysis of Ref. \cite{RG} shows that such fusion rules are indeed
obeyed by the powers of the local dissipation rate in the model
\cite{Kraich1}, and all these operators are dangerous.}
The OPE and the concept of dangerous
operators in the stochastic hydrodynamics were introduced and investigated
in detail in \cite{LOMI,JETP}; see also the review paper \cite{UFN} and
the book \cite{turbo}.

The part of the formal expansion parameter in the RG approach is
played by the exponent $\zeta$ entering into the velocity correlator;
see Eq. (\ref{RC1}) in Sec. \ref{sec:FT} (in Ref. \cite{RG}, it
was denoted by $\eps$, in
order to emphasize the analogy with Wilson's $\eps$ expansion).
The anomalous exponents were calculated in \cite{RG} to the order
$\zeta^{2}$ of the expansion in $\zeta$ for any space dimension,
and they are in agreement with the first-order results obtained
within the zero-mode approach in \cite{Falk1,Falk2,GK,BGK}. In
\cite{RG1}, the RG method was generalized to the case of a
nonsolenoidal (``compressible'') velocity field.

The main advantage of the RG approach (apart from its calculational
efficiency) is the universality: it is not
related to the aforementioned solvability of the rapid-change
model and can equally be applied to the case of finite correlation time,
provided the corresponding model possesses the RG symmetry.
In \cite{RG}, the results were presented for the opposite limiting case
of the time-independent (``frozen'') velocity field.

In this paper, we apply the RG and OPE technique to the problem of a
passive scalar field advected by a self-similar synthetic Gaussian
velocity field with finite correlation time; the steady state is
maintained by an imposed linear mean gradient.
The velocity field satisfies a linear stochastic equation with
effective viscosity and stirring force. The model was proposed and
studied in detail
(using numerical simulations, in two dimensions) in \cite{synth};
its rapid-change version is discussed in \cite{Pumir,SSP,BFL,Pumir2}.
We consider the problem in an arbitrary space dimension, $d\ge2$; we
also stress that the correlation time is not supposed to be small.
We establish the existence in the
inertial-convective range of several different scaling regimes and
show that for some of them the structure functions and other
correlation functions of the problem exhibit anomalous scaling
behavior; we derive explicit analytical expressions for the
corresponding anomalous exponents.

The advection of a passive scalar field in the presence of an
imposed linear gradient is described by the equation
\begin{equation}
\nabla _t\theta=\nu _0\partial^{2} \theta-\h\cdot{\bf v} , \quad
\nabla _t\equiv \partial _t+ v_{i}\, \partial_{i}.
\label{1}
\end{equation}
Here $\theta(x)\equiv \theta(t,{\bf x})$ is the random (fluctuation)
part of the total scalar field $\Theta(x)=\theta(x)+\h\cdot{\bf x}$,
$\h$ is a constant vector that determines distinguished direction,
$\nu _0$ is the molecular diffusivity coefficient,
$\partial _t \equiv \partial /\partial t$,
$\partial _i \equiv \partial /\partial x_{i}$,
$\partial^{2}\equiv\partial _i\partial _i$
is the Laplace operator, and ${\bf v}(x)=\{v_i(x)\}$ is the transverse
(owing to the incompressibility) velocity field.

The velocity obeys the linear stochastic equation, cf. \cite{synth}
\begin{equation}
\partial_{t} v_i + R v_i =f_{i},
\label{NS}
\end{equation}
where $R$ [in the momentum representation $R=R(k)$] is a linear
operation to be specified below and $f_{i}$ is
an external random stirring force with zero mean and the correlator
\begin{equation}
\langle f_{i}(x) f_{j}(x')\rangle =
\int \frac{d\omega}{2\pi}
\int \frac{d{\bf k}}{(2\pi)^d}  P_{ij}({\bf k})\, D^{f}(\omega,k)
\exp [ -{\rm i} (t-t')+{\rm i}{\bf k}\cdot({\bf x}-{\bf x'})] .
\label{f}
\end{equation}
Here $P_{ij}({\bf k}) = \delta _{ij} - k_i k_j / k^2$ is the transverse
projector, $k\equiv |{\bf k}|$ is the wavenumber, and $d$ is the
dimensionality of the ${\bf x}$ space. Following \cite{synth}, we
choose the correlator $D^{f}$ to be independent of the frequency,
so that Eq. (\ref{f}) contains the delta-function in time. More specific,
we choose
\begin{equation}
D^{f}(\omega,k)= g_{0}\nu_0^{3}\, \sigma_{k}^{4-d-\eps-\eta},
\quad R(k)=u_{0}\nu_0\, \sigma_{k}^{2-\eta},
\label{Fin1}
\end{equation}
where
\begin{equation}
\sigma_{k}\equiv \sqrt {k^{2}+m^{2}}.
\label{Fin4}
\end{equation}
The positive amplitude factors $g_{0}$ (a formal small parameter of the
ordinary perturbation theory) and $u_{0}$ are the analogs of the
the coupling constant (``charge'')  $\lambda_{0}$ in the standard
$\lambda_{0}\phi^{4}$ model of critical behavior, see, e.g.,
\cite{Zinn,book3};
in what follows we shall also term these parameters ``coupling constants.''
The exponents $\eps$ and $\eta$ are the analogs of the RG expansion
parameter $\eps=4-d$ in the $\lambda_{0}\phi^{4}$ model,
and we shall use the traditional term  ``$\eps$  expansion'' in our model
for the double expansion
in the $\eps$--$\eta$ plane around the origin $\eps=\eta=0$, with the
additional convention that $\eps=O(\eta)$.
The infrared (IR)
regularization is provided by the integral scale $L\equiv 1/m$; its
precise form is not essential. For $k>>m$ the functions
(\ref{Fin1}) take on simple powerlike form. Dimensionality
considerations show that the charges are related to the characteristic
ultraviolet (UV) momentum scale $\Lambda$ by
\begin{equation}
g_{0}\simeq \Lambda^{\eps+\eta},\quad u_{0}\simeq \Lambda^{\eta}.
\label{gg}
\end{equation}

From Eqs. (\ref{NS}) and (\ref{f}) it follows that ${\bf v}(x)$ obeys
Gaussian distribution with zero mean and correlator
(dropping the transverse projector)
\begin{equation}
D_{v}(\omega,k)=  \frac{D^{f}(k)}{\omega^{2}+R^{2}(k)}
=\frac{g_{0}\nu_0^{3}\, \sigma_{k}^{4-d-\eps-\eta}}
{\omega^{2}+[u_{0}\nu_0\, \sigma_{k}^{2-\eta}]^{2}}\, .
\label{Fin}
\end{equation}
Therefore, the exponent $\eps$ describes the inertial-range behavior of
the equal-time velocity correlator or, equivalently, the energy spectrum
\begin{equation}
E(k) \simeq k^{d-1} \int d\omega D_{v}(\omega,k)
\simeq (g_{0}\nu_0^{2}/u_{0})\, k^{1-\eps},
\label{spectrum}
\end{equation}
cf. \cite{AvelMaj,AvelMaj2,Glimm}, where a close family of models for the
velocity field has been considered for a strongly anisotropic shear flow.
The second exponent, $\eta$, is related to the function
$R(k)$, the reciprocal of the correlation time at the wavenumber $k$
($\eta\equiv2-z$ in the notation of \cite{AvelMaj,AvelMaj2,Glimm,Falk3,Eyink};
our exponents are defined so that $\eps=\eta=0$ correspond to the starting
point of the RG expansion). It then follows that $\eps=8/3$ gives the
Kolmogorow ``five-thirds law'' for the spatial velocity statistics, and
$\eta=4/3$ corresponds to the Kolmogorov frequency.

It was pointed out in \cite{synth} that the linear model (\ref{NS})
suffers from the lack of Galilean invariance and therefore does not take
into account the self-advection of turbulent eddies.
It is well known that the different-time correlations of the Eulerian
velocity field are not self-similar, as a result of these
``sweeping effects,'' and depend substantially on the integral scale;
see, e.g., \cite{sweep}.
Nevertheless, the results of \cite{synth} show that the model gives
reasonable description of the passive advection in an appropriate frame,
where the mean velocity field vanishes. To justify the model
(\ref{NS}), we also note that we shall be interested preferably in the
equal-time, Galilean invariant quantities (structure functions,
correlations of the dissipation rate etc.), which are not affected by
the sweeping effects, and we expect that their absence from the model
(\ref{NS}) is not essential.

We also note that the model contains two special cases
that possess some interest on their own.
In the limit $u_{0}\to\infty$, $g_{0}'\equiv g_{0}/u_{0}^{2}=\const$
we arrive at the rapid-change model:
\begin{equation}
D_{v}(\omega,k)\to g_{0}'\nu_0\,(k^{2}+m^{2})^{-d/2-\zeta/2},
\quad \zeta\equiv \eps - \eta,
\label{RC1}
\end{equation}
and the limit $u_{0}\to 0$, $g_{0}''\equiv g_{0}/u_{0}=\const$
corresponds to the case of a frozen velocity field:
\begin{equation}
D_{v}(\omega,k)\to g_{0}''\nu_0^{2}\,(k^{2}+m^{2})^
{-d/2+1-\eps/2}\, \pi\,\delta(\omega),
\label{RC2}
\end{equation}
when the velocity correlator is independent of the time variable
$t-t'$ in the $t$ representation. The latter case for $\h=0$ has
a close formal resemblance with the well-known models of the random
walks in random environment with long-range correlations; see
\cite{walks1,walks3}.

In Sec. \ref{sec:FT}, we give the field theoretic formulation
of the problem and discuss some its consequences; we also explain briefly
why the ordinary perturbation theory fails to give correct IR
behavior for some values of $\eps$ and $\eta$ and establish the
relationship between the IR and UV problems. In Sec. \ref{sec:RG},
we discuss the UV renormalization of the model, derive the RG
equations and present the one-loop expressions for the basic RG
functions (beta functions and anomalous dimensions). In Sec.
\ref{sec:Fixed}, the analysis of the scaling behavior is given.
Depending on the values of the exponents $\eps$ and $\eta$ entering
into the velocity correlator, the model exhibits various types of
IR scaling regimes, associated with the IR stable fixed points of
the RG equations:

(i) The anomalous scaling behavior with universal (in the
above sense) exponents, characteristic of the rapid-change model,
takes place for $\eta<\eps<2\eta$. The anomalous exponents
depend on the only exponent $\zeta$ entering into Eq. (\ref{RC1}).

(ii) The anomalous scaling behavior with the universal exponents,
characteristic of the model with time-independent (frozen)
velocity field, emerges in the region $\eps>0$, $\eps >2\eta$.
The exponents are determined solely by the equal-time velocity
correlator and depend on the only exponent $\eps$ entering into
Eq. (\ref{RC2}).

(iii) The intermediate regime with nonuniversal exponents, which
depend on the amplitudes entering into the velocity correlator,
emerges for $\eps =2\eta$; the Kolmogorov-type synthetic velocity
field \cite{synth} and the case of a local turnover exponent
\cite{Falk3} correspond to this regime. The nonuniversality of
the exponents in this regime is in agreement with the findings of
Ref. \cite{Falk3}, where the large $d$ limit has been considered.
[However, the exponents in our model turn out to be universal in the
one-loop approximation].

(iv) The diffusive-type regimes, for which the advection (i.e., the
nonlinearity in Eq. (\ref{1})) can be treated within the ordinary
perturbation theory. These regimes take place in the region specified
by the inequalities $\eta>0$, $\eta>\eps$ and $\eta<0$, $\eps<0$.

To avoid possible misunderstandings we emphasize that the limits $g_{0}$,
$u_{0}\to0$ or $g_{0}$, $u_{0}\to\infty$ are not supposed to be
performed in the original correlation function (\ref{Fin}); the parameters
$g_{0}$, $u_{0}$ are fixed at some finite values. The behavior specific
to the models (\ref{RC1}), (\ref{RC2}) arises asymptotically in the
regimes (i) and (ii) as a result of the solution of the RG equations,
when the ``RG flow'' approaches the corresponding fixed point.
Therefore, we deal with the finite correlation time, and there is no
problem with the steady state in the frozen case even in two dimensions.
The regions of IR stability of the regimes (i)--(iv) in the
$\eps$--$\eta$ plane, given above, are identified to the first order
of the $\eps$ expansion, but some of their boundaries are found exactly.

In the regimes (i)--(iii), the correlation functions of the model exhibit
anomalous scaling behavior, i.e., singular dependence on the IR scale $m$
with nonlinear ``anomalous exponents.'' Within the RG and OPE approach,
the latter are related to the scaling dimensions of the tensor
composite operators $\partial\theta\cdots\partial\theta$; these
dimensions are calculated explicitly to the first order of the $\eps$
expansion (one-loop approximation) in Sec. \ref{sec:Operators}.
The inertial-convective-range asymptotic expressions for the structure
functions of arbitrary order (even
and odd) and the equal-time correlations of the scalar gradients
are obtained in Sec. \ref{sec:OPE} using the OPE.

As the exponents $\eps$ and $\eta$ increase, the powers of the velocity
field also become dangerous, and their contributions to the OPE should
be summed. The required summation is performed in Sec. \ref{sec:summ}
on the example of the second-order structure function in the ``frozen''
regime; for the rapid-change regime the problem is absent. This
summation might be interesting as a possible model of the origin of
the anomalous scaling in the structure functions of the velocity itself:
it was argued in \cite{AV} that the singular $m$ dependence of the
equal-time correlators for the stochastic Navier--Stokes (NS) equation
is related to {\it infinite} families of dangerous operators.

The formulation (\ref{gg}) is typical to the models of critical behavior
[i.e., the dimensional coupling constants are expressed only
through the UV scale], and we shall call it ``standard.'' In
\cite{synth,Pumir2,ShS,ShS2,PDF,Falk3,Eyink}, a different version of
the problem was considered, in which the velocity correlator contains
nontrivial dependence on the integral turbulence scale. This ``exotic''
(from the viewpoints of the theory of critical behavior) formulation
requires special attention; it is considered briefly in Sec. \ref{sec:Exo}.

The results obtained are reviewed in Sec. \ref{sec:Con}, where we also
discuss briefly the passive advection by the non-Gaussian velocity field
governed by the nonlinear stochastic NS equation. Our approach is
generalized directly to this case, and the explicit expressions for
the anomalous exponents can readily be obtained in the first order
of the corresponding $\eps$ expansion. We also discuss new problems
that arise in the NS model beyond the $\eps$ expansion.

 \section{Field theoretic formulation of the model.
 IR and UV singularities in the perturbation theory} \label {sec:FT}

According to the general theorem (see, e.g., Refs. \cite{Zinn,book3}),
the stochastic problem (\ref{1})--(\ref{f}) is equivalent
to the field theoretic model of the doubled set of fields
$\Phi\equiv\{ \theta, \theta',{\bf v}, {\bf v'}\}$ with action
functional
\begin{equation}
S(\Phi)= (1/2) {\bf v'} D^{f} {\bf v'} +
{\bf v'} [- \partial_{t}{\bf v}- R{\bf v}]+
\theta' \left[ - \partial_{t}\theta -({\bf v}\partt) \theta
+ \nu _0\partial^{2} \theta - \h\cdot{\bf v}\right].
\label{action1}
\end{equation}
Here $D^{f}$ is the correlator (\ref{Fin1}), the required integrations
over $x=(t,{\bf x})$ and summations over the vector indices in Eq.
(\ref{action1}) and analogous formulas below are implied.

The formulation (\ref{action1}) means that statistical averages
of random quantities in the stochastic problem (\ref{1})--(\ref{f})
coincide with functional averages with the weight $\exp S(\Phi)$,
so that generating functionals of total [$G(A)$] and connected
[$W(A)$] Green functions are represented by the functional integral
\begin{equation}
G(A)=\exp  W(A)=\int {\cal D}\Phi \exp [S(\Phi )+A\Phi ]
\label{gene}
\end{equation}
with arbitrary sources $A(x)$ in the linear form
\begin{equation}
A\Phi \equiv
\int dx[A^{\theta}(x)\theta (x)+A^{\theta '}(x)\theta '(x)
+ A^{\bf v}_{i}(x)v_{i}(x)+ A^{\bf v'}_{i}(x)v_{i}'(x)].
\label{sour}
\end{equation}
In the following, we shall not be interested in the Green functions
involving the auxiliary vector field ${\bf v'}$, so that we can set
$A^{\bf v'}=0$ in Eq. (\ref{sour}). It is then convenient to perform
the Gaussian integration over ${\bf v'}$ in Eq. (\ref{gene})
explicitly. We arrive at the field theoretic model of the reduced
set of fields $\Phi\equiv\{ \theta, \theta',{\bf v}\}$ with the action
\begin{equation}
S(\Phi)= \theta' \left[ - \partial_{t}\theta -({\bf v}\partt) \theta
+ \nu _0\partial^{2} \theta - \h\cdot{\bf v}\right]
-{\bf v} D_{v}^{-1} {\bf v}/2.
\label{action}
\end{equation}
The first four terms in Eq. (\ref{action}) represent
the Martin--Siggia--Rose-type action
for the stochastic problem (\ref{1}) at fixed ${\bf v}$,
and the last term represents the Gaussian averaging over ${\bf v}$
with the correlator $D_{v}$ from Eq. (\ref{Fin}).

The model (\ref{action}) corresponds to a standard Feynman
diagrammatic technique with the triple vertex
$-\theta'({\bf v}\partt)\theta=\theta'V_{j}v_{j}\theta$
with vertex factor
\begin{equation}
V_{j}= {\rm i} k_{j},
\label{vertex}
\end{equation}
where ${\bf k}$  is the momentum flowing into the vertex via
the field $\theta'$, and the bare propagators
(in the momentum-frequency representation)
\begin{eqnarray}
\langle \theta \theta' \rangle _0=\langle \theta' \theta \rangle _0^*=
(-{\rm i}\omega +\nu _0 k^2)^{-1} ,
\nonumber \\
\langle \theta \theta \rangle _0= \langle \theta \theta' \rangle _0
h_{i}h_{j} \langle v_{i} v_{j}  \rangle _0
\langle \theta' \theta \rangle _0,
\nonumber \\
\langle \theta v_{i} \rangle _0= - \langle \theta \theta' \rangle _0
h_{j} \langle v_{j} v_{i}  \rangle _0,
\nonumber \\
 \langle \theta '\theta '\rangle _0=0 ,
\label{lines}
\end{eqnarray}
where $h_{i}$ is a component of the vector $\h$  and the bare
propagator $\langle v_{i} v_{j}  \rangle _0$
is given by Eq. (\ref{Fin}).

The magnitude $h\equiv|\h|$
can be eliminated from the action (\ref{action}) by rescaling
of the scalar fields: $\theta\to h\theta$, $\theta'\to \theta'/h$.
Therefore,  any total or connected Green function
of the form $\langle\theta(x_{1})\cdots\theta(x_{n})\, \theta'(y_{1})
\cdots\theta'(y_{p})\rangle$ contains the factor of $h^{n-p}$.
The parameter $h$ appears in the bare propagators (\ref{lines})
only in the numerators. It then follows that the Green functions
with $n-p<0$ vanish identically. On the contrary, the 1-irreducible
function $\langle\theta(x_{1})\cdots\theta(x_{n})\, \theta'(y_{1})
\cdots\theta'(y_{p})\rangle_{\rm 1-ir}$ contains a factor of
$h^{p-n}$ and therefore vanishes for $n-p>0$; this
fact will be relevant in the analysis of the renormalizability of
the model (see below).

Another important  consequence of the representation (\ref{gene}),
(\ref{action}) is that the large-scale anisotropy persists, through
the dependence on $\h$, for all ranges of momenta (including
convective and dissipative ranges), and that the dimensionless ratios
of the structure functions are strictly independent on $h$; cf.
\cite{synth,pass1,pass2,ShS,ShS2,PDF}. It is noteworthy that all
these statements equally hold for any statistics of the velocity
field (not necessarily Gaussian or synthetic), provided its
distribution is independent of $\h$.

However, the ordinary perturbation theory fails to give correct
IR behavior of Green functions for some values of the exponents
$\eps$ and $\eta$. This can easily be illustrated on the simplest
example of the 1-irreducible Green function
$\langle\theta'\theta\rangle_{\rm 1-ir}$. It
satisfies the Dyson equation of the form
\begin{equation}
\langle\theta'\theta\rangle_{\rm 1-ir} = -{\rm i} \omega +
\nu_0 k^{2} -\Sigma_{\theta'\theta} (\omega, k),
\label{Dyson}
\end{equation}
where $\Sigma_{\theta'\theta}$ is the self-energy operator
represented by the corresponding 1-irreducible diagrams.
Its one-loop approximation has the form
\begin{equation}
\Sigma_{\theta'\theta} = \put(0.00,-56.00){\makebox{\dS}}
\hskip1.7cm .
\label{Dyson2}
\end{equation}
Here and below the solid lines in the diagrams denote the bare
propagator $\langle\theta\theta'\rangle_{0}$ from Eq. (\ref{lines}),
the end with a slash corresponds to the field $\theta'$, and the
end without a slash corresponds to $\theta$; the dashed lines
denote the bare propagator (\ref{Fin}); the vertices correspond
to the factor (\ref{vertex}). The analytic
expression for the diagram in (\ref{Dyson2}) has the form
\begin{equation}
\Sigma_{\theta'\theta} (\omega, k) = - k_{i}k_{j}
\int \frac{d\omega'}{2\pi} \int \frac{d{\bf q}}{(2\pi)^{d}} \,
\frac{P_{ij}({\bf q})\,D_{v}(\omega',q)} {-{\rm i} (\omega+\omega')
+ \nu_{0} ({\bf q}+{\bf k})^{2}}\, ,
\label{novaja1}
\end{equation}
where $q\equiv|{\bf q}|$ and $D(\omega',q)$ is given by Eq. (\ref{Fin});
the factor of $k_{i}k_{j}$ arises from the vertex factors (\ref{vertex}).
Integration over $\omega'$ in Eq. (\ref{novaja1}) yields
\begin{equation}
\Sigma_{\theta'\theta} (\omega, k) = - k_{i}k_{j} \frac{g_0\nu_0^{2}}
{2u_{0}} \int \frac{d{\bf q}}{(2\pi)^{d}} \,
\frac{P_{ij}({\bf q})\, \sigma_{q}^{2-d-\eps}}
{-{\rm i} \omega +\nu_0 ({\bf q}+{\bf k})^{2} + u_{0}\nu_0
\sigma_{q}^{2-\eta}}\, .
\label{novaja2}
\end{equation}

We are interested in the IR behavior of the function (\ref{novaja2}),
i.e., the behavior of small $k$, $\omega$ and $m$. It is easily seen
that this behavior is nontrivial in the region on the $\eps$--$\eta$ plane,
determined by the inequalities $\eta<0$, $\eps>0$ and $\eta>0$, $\eps>\eta$,
because the integral in (\ref{novaja2})
is then IR divergent if $k$, $\omega$ and $m$ are simply set equal to zero.
On the contrary, for the rest of the $\eps$--$\eta$ plane,
the leading term
of the desired asymptotic behavior is indeed obtained simply by
setting $k=\omega=m=0$. The analysis is extended directly to the
higher-order diagrams; it shows that these IR singularities enhance as
the order of a diagram increases, and that they take place only within
the same region on the $\eps$--$\eta$ plane. The IR singularities
compensate the smallness of the coupling constant $g_{0}$, assumed
within the framework of the ordinary perturbation theory. Therefore,
in order to find correct IR behavior we would have to sum the entire
series even if the expansion parameter, $g_{0}$, were small.

It is also clear that these IR singularities get weaker as the
parameters $\eps$, $\eta$ decrease, and they would disappear at
$\eps=\eta=0$ if we could take this limit in Eq. (\ref{novaja2}).
However, this is impossible owing to the UV divergence in the
integral (\ref{novaja2}) at this point. In general, the diagrams
of $\Sigma_{\theta'\theta}$ are UV divergent in the region
$\eta>0$, $\eps<0$ and $\eta<0$, $\eps<\eta$, and the UV cutoff
at $q\equiv|{\bf q}|\simeq\Lambda$
is then implied in (\ref{novaja2}) and higher-order diagrams.
If the point $\eps=\eta=0$ is approached from inside the region of
UV convergence, the UV singularities manifest themselves as poles in
$\eps$, $\eta$ and their linear combinations. The elimination of these
poles is the classical UV problem, and its solution is given by
the standard theory of UV renormalization; the RG equations are obtained
within the framework of this theory and express the simple idea of
nonuniqueness of the renormalization procedure. The correlation
between the IR and UV singularities near the ``logarithmic point''
$\eps=\eta=0$, noted above, explains to some extent why the RG method,
which is closely related to the UV divergences, can be a useful
tool in studying the IR behavior, and why the exponents $\eps$
and $\eta$ are expected to be relevant small parameters
in the RG expansions.

Surprisingly, simple arguments given above lead to
reasonable conclusions: the rigorous RG analysis confirms that
the Green functions of the model indeed show anomalous IR behavior
for some values of $\eps$ and $\eta$, and the region determined by
the inequalities $\eta<0$, $\eps>0$ and $\eta>0$, $\eps>\eta$
coincides with the region of stability of the corresponding fixed points
in the {\it linear} approximation; see Sec. \ref{sec:RG}, \ref{sec:Fixed}.

 \section{UV renormalization of the model. RG functions and RG equations}
 \label {sec:RG}

The renormalization of the model (\ref{action}) is similar to the
renormalization of the simpler rapid-change model, considered in
detail in \cite{RG}; below we confine ourselves to only the
necessary information.

The analysis of UV divergences is based on the analysis of {\it canonical
dimensions}, see \cite{book3,Collins}. Dynamical models of the type
(\ref{action}), in contrast to static models, are two-scale
\cite{UFN,turbo,Pismak}, i.e., the action functional (\ref{action}) is
invariant with respect to the two independent scale transformations,
$S(\Phi', z_{i}')=S(\Phi, z_{i})$, where $\Phi\equiv\{ \theta,
\theta',{\bf v}\}$ and $z_{i}=\{g_0,u_{0},\nu_0,m\}$
is the full set of the model parameters. In the first transformation,
the time variable is fixed and the space variable is dilated along with
all the fields and parameters:
\begin{equation}
\Phi(t,{\bf x})\to \Phi'(t,{\bf x})=
\lambda^{d_{\Phi}^{k}} \Phi(t,\lambda{\bf x}),
\quad z_{i}\to  z_{i}' = \lambda^{d_{z_{i}}^{k}} z_{i},
\label{scale1}
\end{equation}
and in the second the space variable is fixed and all the other
quantities are dilated:
\begin{equation}
\Phi(t,{\bf x})\to
\Phi'(t,{\bf x})=  \lambda^{d_{\Phi}^{\omega}} \Phi(\lambda t,{\bf x}),
\quad z_{i}\to  z_{i}' = \lambda^{d_{z_{i}}^{\omega}} z_{i}.
\label{scale2}
\end{equation}
Here $\lambda>0$ is an arbitrary transformation parameter, and two
independent canonical dimensions, the momentum dimension $d_{F}^{k}$ and
the frequency dimension $d_{F}^{\omega}$, are assigned to each quantity
$F$ (a field or a parameter in the action functional). These canonical
(``engineering'') dimensions should not be confused with the exact
critical dimensions: the latter are subject to nontrivial calculation,
while the former are simply determined from the natural normalization
conditions $d_k^k=-d_{\bf x}^k=1$, $d_k^{\omega}=d_{\bf x}^{\omega}=0$,
$d_{\omega }^k=d_t^k=0$, $d_{\omega}^{\omega}=-d_t^{\omega}=1$,
and from the requirement that each term of the action functional be
dimensionless [i.e., be invariant with respect to the transformations
(\ref{scale1}) and (\ref{scale2}) separately]. Then, based on $d_{F}^{k}$
and $d_{F}^{\omega}$, one can introduce the total canonical dimension
\cite{UFN,turbo,Pismak}, which corresponds to the dilatation with fixed
value of $\nu_0$ (i.e., zero canonical dimension can be assigned to
$\nu_0$). In our model, $\partial_{t}\propto\nu_0\partial^{2}$, so that
the total dimension is given by $d_{F}=d_{F}^{k}+2d_{F}^{\omega}$.

In the action (\ref{action}), there are fewer terms than fields
and parameters, and the canonical dimensions are not determined
unambiguously. This is of course a manifestation of the fact that
the ``superfluous'' parameter $h=|\h|$ can be eliminated from the
action; see above. After it has been eliminated
(or, equivalently, zero canonical dimensions have been assigned
to it), the definite canonical dimensions can be assigned to the
other quantities. They are given in Table \ref{table1},
including the dimensions of renormalized parameters,
which will appear later on.

From Table \ref{table1} it follows that the model is logarithmic
(the both coupling constants $g_{0}$ and $u_{0}$ are dimensionless)
at $\eps=\eta=0$. This means that the UV divergences
in the Green functions have the form of the poles in $\eps$,
$\eta$, and all their possible linear combinations.

The total dimension $d_{F}$ plays in the theory of
renormalization of dynamical models the same role as does
the conventional (momentum) dimension in static problems.
The canonical dimensions of an arbitrary
1-irreducible Green function $\Gamma = \langle\Phi \dots \Phi
\rangle _{\rm 1-ir}$ are given by the relations
\begin{equation}
d_{\Gamma }^k=d- N_{\Phi}d_{\Phi}^k, \qquad
d_{\Gamma }^{\omega }=1-N_{\Phi }d_{\Phi }^{\omega }, \qquad
d_{\Gamma }=d_{\Gamma }^k+2d_{\Gamma }^{\omega }=
d+2-N_{\Phi }d_{\Phi},
\label{17}
\end{equation}
where $N_{\Phi}=\{N_{\theta},N_{\theta'},N_{\bf v}\}$ are the
numbers of corresponding fields entering into the function
$\Gamma$, and the summation over all types of the fields is
implied.
The total dimension $d_{\Gamma}$ is the formal index of the
UV divergence. Superficial UV divergences, whose removal requires
counterterms, can be present only in those functions $\Gamma$ for
which $d_{\Gamma}$ is a non-negative integer.

Analysis of divergences in the problem (\ref{action}) should be based on
the following auxiliary considerations; cf. \cite{RG,UFN,turbo}:

(1) All the 1-irreducible Green functions with
$N_{\theta'}< N_{\theta}$ vanish; see Sec. \ref{sec:FT}.

(2) If for some reason a  number of external momenta occur as an
overall factor in all the diagrams of a given Green function, the
real index of divergence $d_{\Gamma}'$ is smaller than $d_{\Gamma}$
by the corresponding number (the Green function requires
counterterms only if $d_{\Gamma}'$  is a non-negative integer).

In the model (\ref{action}), the derivative $\partial$ at the
vertex $\theta'({\bf v}\partt)\theta$ can be moved onto the
field $\theta'$ by virtue of the transversality of the field
${\bf v}$. Therefore, in any 1-irreducible diagram it is always
possible to move the derivative onto any of the external
``tails'' $\theta$ or $\theta'$, which decreases the real index
of divergence: $d_{\Gamma}' = d_{\Gamma}- N_{\theta}-N_{\theta'}$.
This also means that the fields $\theta$, $\theta'$ enter into
the counterterms only in the form of the derivatives
$\partial\theta$ and $\partial\theta'$.

From the dimensions in Table \ref{table1} we find
$d_{\Gamma} = d+2 - N_{\bf v}
+ N_{\theta}- (d+1)N_{\theta'}$  and $d_{\Gamma}'=(d+2)(1-N_{\theta'})
- N_{\bf v}$. From these expressions it follows that for any $ d$,
superficial divergences can exist only in the 1-irreducible functions
$\langle\theta'\theta\dots\theta\rangle_{\rm 1-ir}$ with $N_{\theta'}=1$
and arbitrary value of $N_{\theta}$, for which $d_{\Gamma}=1+N_{\theta}$,
$d_{\Gamma}'=0$. However, all the functions with $N_{\theta}>
N_{\theta'}$ vanish (see above) and obviously do not require
counterterms. As in the case of the rapid-change model
\cite{RG,RG1}, we are left with the only superficially
divergent function $\langle\theta'\theta\rangle_{\rm 1-ir}$;
the corresponding counterterm contains two symbols
$\partial$  and is therefore reduced to $\theta'\partial^{2}\theta$.
The inclusion of this counterterm is reproduced by the multiplicative
renormalization of the parameters $g_{0}$, $u_{0}$, and $\nu_0$
in the action functional (\ref{action}):
\begin{equation}
\nu_{0}=\nu Z_{\nu}, \qquad g_{0}=g\mu^{\eps+\eta}\, Z_{g},
\qquad u_{0}=u\mu^{\eta}\, Z_{u},
\label{mult}
\end{equation}
where the dimensionless parameters $g$, $u$, and $\nu$ are
the renormalized analogs of the
bare parameters, $\mu$ is the renormalization mass in the
minimal subtraction (MS) scheme, which we always use in
practical calculations, and $Z_{i}=Z_{i}(g,u)$ are the
renormalization constants. They satisfy the identities
\begin{equation}
Z_{g}= Z_{\nu}^{-3},\quad Z_{u}= Z_{\nu}^{-1},
\label{svaz}
\end{equation}
which result from the absence of the renormalization
of the contribution with $D_{v}$ in the functional (\ref{action}).
No renormalization of the fields,  the ``mass'' $m$, and the vector $\h$
is required, i.e., $Z_{\Phi}=1$ for all $\Phi$ and $Z_{m}=Z_{h}=1$.

The renormalized action functional has the form
\begin{equation}
S_{\rm ren}(\Phi)=
\theta' \left[ - \partial_{t}\theta -({\bf v}\partt) \theta
+ \nu Z_{\nu}\partial^{2} \theta -\h\cdot{\bf v}\right]
-{\bf v} D_{v}^{-1} {\bf v}/2,
\label{actionR}
\end{equation}
where the correlator $D_{v}$ is expressed in renormalized
parameters using the formulas (\ref{mult}):
\begin{equation}
D_{v}(\omega,k)= \frac{g\nu^{3}\mu^{\eps+\eta}
\sigma_{k}^{4-d-\eps-\eta}}
{\omega^{2}+[u\nu\mu^\eta \sigma_{k}^{2-\eta}]^{2}}\, .
\label{FinR}
\end{equation}

The relation $ S(\Phi,e_{0})=S_{\rm ren}(\Phi,e,\mu)$ (where $e_{0}$
is the complete set of bare parameters, and $e$ is the set of renormalized
parameters) for the generating functional $W(A)$ in Eq. (\ref{gene})
yields $ W(A,e_{0})=W_{\rm ren}(A,e,\mu)$. We use $\widetilde{\cal D}_{\mu}$
to denote the differential operation $\mu\partial_{\mu}$ for fixed
$e_{0}$ and operate on both sides of this equation with it. This
gives the basic RG differential equation:
\begin{equation}
{\cal D}_{RG}\,W_{\rm ren}(A,e,\mu) = 0,
\label{RG1}
\end{equation}
where ${\cal D}_{RG}$ is the operation $\widetilde{\cal D}_{\mu}$
expressed in the renormalized variables:
\begin{equation}
{\cal D}_{RG}\equiv {\cal D}_{\mu} + \beta_{g}(g,u)\partial_{g} +
\beta_{u}(g,u)\partial_{u} -\gamma_{\nu}(g,u){\cal D}_{\nu}.
\label{RG2}
\end{equation}
In Eq. (\ref{RG2}), we have written ${\cal D}_{x}\equiv x\partial_{x}$ for
any variable $x$, and the RG functions (the $\beta$ functions and
the anomalous dimension $\gamma$) are defined as
\begin{mathletters}
\label{RGF}
\begin{equation}
\gamma_{\nu}\equiv \Dm \ln Z_{\nu},
\label{RGF1}
\end{equation}
\begin{equation}
\beta_{g}\equiv\Dm  g=g[-\eps-\eta+3\gamma_{\nu}],
\label{beta1}
\end{equation}
\begin{equation}
\beta_{u}\equiv\Dm  u=u[-\eta+\gamma_{\nu}].
\label{beta2}
\end{equation}
\end{mathletters}
The relations between $\beta$ and $\gamma$ in Eq. (\ref{RGF})
result from the definitions and the relation (\ref{svaz}).

Now let us turn to the explicit calculation of the constant $Z_{\nu}$
in the one-loop approximation in the MS scheme.
The constant $Z_{\nu}$ is determined by the requirement that
the 1-irreducible Green function
$\langle\theta'\theta\rangle_{\rm 1-ir}$,
when expressed in renormalized variables, be UV finite [i.e.,
have no singularities for $\eps$, $\eta\to0$].
The Dyson equation (\ref{Dyson}) relates this function to the
self-energy operator $\Sigma_{\theta'\theta}$, and Eq. (\ref{novaja2})
gives the explicit expression for the latter in the first order
$O(g_{0})$ of the unrenormalized perturbation theory. Now
we have to calculate the function $\Sigma_{\theta'\theta}$ in the
order $O(g)$ of the renormalized perturbation theory; therefore
we should simply replace $\nu_{0}\to\nu$ in the propagator
$\langle\theta\theta'\rangle_{0}$ and use the expression
(\ref{FinR}) for the velocity correlator in Eq. (\ref{Dyson2}), which
leads to the substitution $g_{0}\to g\mu^{\eps+\eta}$,
$u_{0}\to u \mu^{\eta}$, $\nu_{0}\to\nu$ in Eqs. (\ref{novaja1}),
(\ref{novaja2}). We know that the
divergent part of the diagram is independent of $\omega$,
so that we can set $\omega=0$ in what follows.
It is also convenient to cut off the integral over ${\bf q}$ from below
at $q\simeq m$ and set $m=0$ in the integrand (the integral
diverges logarithmically, and its UV divergent part is independent of
the specific form of the IR
regularization). Furthermore, we can set ${\bf k}=0$ in the integrand
(we know that the counterterm is proportional to $k^{2}$, and the factor
of $k^{2}$ has already been isolated from the integral)  and make use
of the isotropy, namely,
\[\int d{\bf q}\, f(q) P_{ij}({\bf q}) = \delta_{ij}\frac{d-1}{d}
\int d{\bf q}\, f(q). \]
Then Eq. (\ref{novaja2}) yields
\begin{equation}
\Sigma_{\theta'\theta} (\omega=0, k) \simeq - k^{2} \,
\frac{g\nu \mu^{\eps} \,(d-1)\, J}{2ud}\, ,
\label{Dyson4}
\end{equation}
where we have written
\begin{equation}
J\equiv \int \frac{d{\bf q}}{(2\pi)^{d}} \,
\frac{q^{-d-\eps}}{1+u\,(\mu/q)^{\eta}}
\label{J}
\end{equation}
and $\simeq$ denotes the equality up to the UV finite parts.
The expansion of the integrand in $u$ gives
\begin{equation}
J=  \sum_{s=0}^{\infty} (-u)^{s}\mu^{s\eta}
\int \frac{d{\bf q}}{(2\pi)^{d}} q^{-d-\eps-s \eta}\simeq
\frac{S_{d}}{(2\pi)^{d}} \sum_{s=0}^{\infty} (-u)^{s}
\frac{\mu^{s\eta}\,m^{-\eps-s\eta}}{\eps+s\eta}\, ,
\label{J1}
\end{equation}
where the parameter $m$ arises from the IR limit in the integral
over ${\bf q}$ and $S_d\equiv 2\pi ^{d/2}/\Gamma (d/2)$
is the surface area of the unit sphere in $d$-dimensional space.

Finally, from Eqs. (\ref{Dyson4}) and (\ref{J1})
we obtain
\begin{equation}
\Sigma_{\theta'\theta} (\omega=0, k) \simeq
\frac{-ag\nu k^2}{u} \sum_{s=0}^{\infty}
\frac{(-u)^{s}\,(\mu/m)^{\eps+s\eta}} {\eps+s\eta},
\label{Dyson5}
\end{equation}
where we have written
\begin{equation}
a\equiv \frac{(d-1)S_{d}}{2d(2\pi)^{d}}.
\label{a}
\end{equation}

The renormalization constant $Z_{\nu}$ is found from the requirement
that the UV divergences cancel out in Eq. (\ref{Dyson}) after the
substitution $\nu_0=\nu Z_{\nu}$. This determines $Z_{\nu}$ up to an
UV finite contribution; the latter is fixed by the choice of the
renormalization scheme. In the MS scheme all the renormalization
constants have the form ``1 + only poles in $\eps$, $\eta$ and their
linear combinations,'' which gives the following expression
\begin{equation}
Z_{\nu}=1- \frac{ag}{u}\sum^{\infty}_{s=0} \frac{(-u)^{s}}
{\eps+s\eta}\, ,
\label{Z}
\end{equation}
with the coefficient $a$ from Eq. (\ref{a}).

In contrast to the rapid-change model, the one-loop
approximation in the case at hand is not exact: the expression
(\ref{Z}) has nontrivial corrections of order $g^{2}$,
$g^{3}$, and so on.  The series in Eq. (\ref{Z}) can be expressed
in the form of a single integral, but this is not convenient for
the calculation of the RG functions.

The RG functions in the one-loop approximation can be calculated
from the re\-nor\-malization constant (\ref{Z})
using the identity $\Dm=\beta_{g}\partial_{g}+\beta_{u}\partial_{u}$,
which follows from the definitions (\ref{RGF})
and the fact that $Z_{\nu}$ depends only on the charges $g,u$.
Within our accuracy this identity is reduced to
$\Dm\simeq  -(\eps+\eta)\D_g -\eta\D_u $.
From Eq. (\ref{Z}) it then follows:
\begin{eqnarray}
\gamma_{\nu}= \biggl[(\eps+\eta)\D_g +\eta\D_u\biggr]
\frac{ag}{u}\sum^{\infty}_{s=0} \frac{(-u)^{s}}
{\eps+s\eta}=
\frac{ag}{u}\sum^{\infty}_{s=0} (-u)^{s}= \frac{ag}{u(1+u)}\, ,
\label{gammanu}
\end{eqnarray}
up to the corrections of order $g^{2}$ and higher. The beta functions
are obtained from Eq. (\ref{gammanu}) using the relations
(\ref{beta1}), (\ref{beta2}).

 \section{Fixed points and scaling regimes}
 \label {sec:Fixed}

It is well known that possible scaling regimes of a renormalizable
model are associated with the IR stable fixed points of the
corresponding RG equations, see, e.g., \cite{Zinn,book3}. The fixed
points are determined from the requirement that all the beta functions
of the model vanish. In our model the coordinates $g_{*},u_{*}$ of
the fixed points are found from the equations
\begin{equation}
\beta_{g} (g_{*},u_{*})=\beta_{u} (g_{*},u_{*})=0
\label{points}
\end{equation}
with the beta functions given in Eqs. (\ref{beta1}), (\ref{beta2}).
The type of the fixed point is determined by the eigenvalues of
the matrix $\Omega=\{\Omega_{ik}=\partial\beta_{i}/\partial g_{i}\}$,
where $\beta_{i}$ denotes the full set of the beta functions and
$g_{i}$ is the full set of charges. For the standard (as in Eq.
(\ref{gg})) formulation of the problem the IR asymptotic behavior
is governed by the IR stable fixed points, i.e., those for
which all the eigenvalues are positive.

From the equations (\ref{beta1}), (\ref{beta2}) we obtain the exact
relation $\beta_{g}/g-3\beta_{u}/u =2\eta-\eps$. It shows that the
beta functions $\beta_{g}$, $\beta_{u}$ cannot vanish simultaneously
for finite values of their arguments. [The only exception is the case
$2\eta=\eps$. We shall study it separately, and for now we assume
$2\eta\ne\eps$.]  Therefore, to find the fixed points we must set
either $u=0$ or $u=\infty$ and simultaneously rescale $g$ so that
the anomalous dimension $\gamma_{\nu}$ remain finite.

In order to study the limit $u\to\infty$ we change to
the new variables $w\equiv 1/u$, $g'\equiv g/u^{2}$;
the corresponding beta functions have the form
\begin{eqnarray}
\beta_{w}\equiv \Dm w= -\beta_{u}/u^{2}=w[\eta-\gamma_{\nu}],
\nonumber \\
\beta_{g'}\equiv \Dm g'=\beta_{g}/u^{2}-2g\beta_{u}/u^{3}=
g'[\eta-\eps+\gamma_{\nu}],
\label{beta'}
\end{eqnarray}
and for the one-loop anomalous dimension we obtain from
Eq. (\ref{gammanu})
\begin{equation}
\gamma_{\nu}=ag'/(1+w)
\label{gamma'}
\end{equation}
with the constant $a$ defined in Eq. (\ref{a}).
From the expressions (\ref{beta'}) we find two fixed points,
which we denote FPI and FPII. The first point is trivial,
\begin{eqnarray}
{\rm FPI:} \qquad w_{*}=g'_{*}=0; \qquad \gamma_{\nu}^{*}=0.
\label{FPI}
\end{eqnarray}
The corresponding matrix $\Omega$ is diagonal with the diagonal
elements
\begin{equation}
\Omega_{1}=\eta, \quad \Omega_{2}=\eta-\eps.
\label{omegaI}
\end{equation}
For the second point we obtain
\begin{eqnarray}
{\rm FPII:} \qquad w_{*}=0,\ g'_{*}= (\eps-\eta)/a; \qquad
\gamma_{\nu}^{*}= \eps-\eta.
\label{FPII}
\end{eqnarray}
The corresponding matrix $\Omega$ is triangular,
$\partial_{g'}\beta_{w}=0$, and its eigenvalues coincide with the
diagonal elements:
\begin{eqnarray}
\Omega_{1}= \partial_{w}\beta_{w} =\eta-\gamma_{\nu}^{*}=2\eta-\eps,
\nonumber \\
\Omega_{2}= \partial_{g'}\beta_{g'} =ag'_{*}= \eps-\eta.
\label{omegaII}
\end{eqnarray}
We note that the expressions for $\gamma_{\nu}^{*}$ in Eq. (\ref{FPII})
and for $\Omega_{1}$ in (\ref{omegaII})  are exact, i.e., they have no
corrections of order $O(\eps^{2})$ [we take $\eps\simeq\eta$, so that
here and below $O(\eps^{2})$
denotes all the terms of the form $\eps\eta$, $\eta^{2}$ and higher].

Now let us turn to the regime with $u\to0$. In order to study this
limit we change to the new variable $g''\equiv g/u$;
the corresponding beta functions have the form
\begin{eqnarray}
\beta_{g''}\equiv\Dm g''=\beta_{g}/u-g\beta_{u}/u^{2}=
g''[-\eps+2\gamma_{\nu}],   \nonumber \\
\beta_{u}=u[-\eta+\gamma_{\nu}]
\label{beta''}
\end{eqnarray}
[the function $\beta_{u}$ is the same as in Eq. (\ref{beta2})].
The one-loop anomalous dimension (\ref{gammanu}) takes the form
\begin{equation}
\gamma_{\nu} = ag''/(1+u) .
\end{equation}
From the expressions (\ref{beta''}) we find two fixed points,
which we denote FPIII and FPIV. The first point is trivial,
\begin{eqnarray}
{\rm FPIII:} \qquad u_{*}=g''_{*}=0; \qquad \gamma_{\nu}^{*}=0.
\label{FPIII}
\end{eqnarray}
The corresponding matrix $\Omega$ is diagonal with the elements
\begin{equation}
\Omega_{1}=-\eps, \quad \Omega_{2}=-\eta.
\label{omegaIII}
\end{equation}
For the nontrivial point we obtain
\begin{eqnarray}
{\rm FPIV:} \qquad u_{*}=0,\ g''_{*}=\eps/2a; \qquad
\gamma_{\nu}^{*}=\eps/2.
\label{FPIV}
\end{eqnarray}
The corresponding matrix $\Omega$ is triangular,
$\partial_{g''}\beta_{u}=0$, and its eigenvalues have the form
\begin{eqnarray}
\Omega_{1}=\partial_{u}\beta_{u} =-\eta+\gamma_{\nu}^{*}=(\eps-2\eta)/2,
\nonumber \\
\Omega_{2}=\partial_{g''}\beta_{g''} = 2ag''_{*}= \eps .
\label{omegaIV}
\end{eqnarray}
The expressions for $\gamma_{\nu}^{*}$ in Eq.
(\ref{FPIV}) and for $\Omega_{1}$ in Eq. (\ref{omegaIV}) are exact.
Of course, the expressions (\ref{omegaI}), (\ref{omegaIII}), and
$\gamma_{\nu}^{*}=0$ for the trivial fixed points are also exact.

In the special case $\eps=2\eta$ the beta functions (\ref{beta1}),
(\ref{beta2}) become proportional, and the set (\ref{points})
reduces to a single equation. As a result, the corresponding
nontrivial fixed point, which we denote FPV, is degenerate:
rather than a point, we have a line of fixed points in the
$g$--$u$ plane. It is given by the relation
\begin{eqnarray}
{\rm FPV:} \qquad g_{*}/u_{*}(u_{*}+1)=\eta/a; \qquad
\gamma_{\nu}^{*}=\eta=\eps/2.
\label{FPV}
\end{eqnarray}
The exact expression for $\gamma_{\nu}^{*}$ follows from the
relation between the RG functions in Eq. (\ref{RGF}). The eigenvalues
of the matrix $\Omega$  (which is not diagonal here) have the form
\begin{equation}
\Omega_{1}=0, \quad \Omega_{2}=\eta\,(2+u_{*})/(1+u_{*}).
\label{omegaV}
\end{equation}
The vanishing of the element $\Omega_{1}$ reflects the existence of
a marginal direction in the $g$--$u$ plane (along the line of the
fixed points) and is therefore an exact fact. The coordinates of a
point on the line (\ref{FPV}) can also be expressed explicitly as
functions of the dimensionless parameter $\rho\equiv g_{0}/u^{3}_{0}$
using the exact relation $g_{0}/u^{3}_{0}=g_{*}/u^{3}_{*}$.
The actual expansion parameter appears to be
$\sqrt\eta$ rather than $\eta$ itself, and the zeroth order
approximation has the form
\begin{equation}
g_{*}=(\eta/a)^{3/2}\rho^{-1/2}, \qquad u_{*}=(\eta/a\rho)^{1/2},
\qquad \Omega_{2}=2\eta.
\label{roots}
\end{equation}

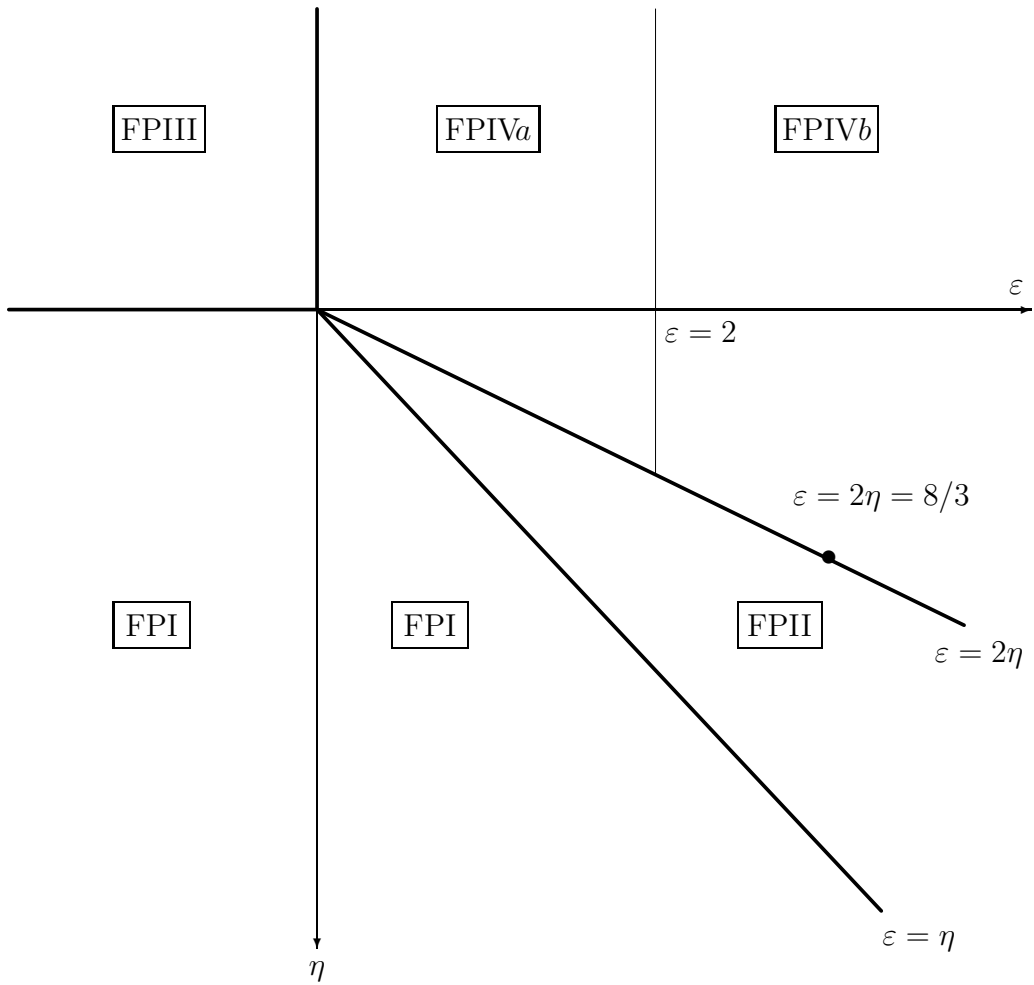
In Figure~I, we show the regions of stability for the fixed points
FPI--FPV in the $\eps$--$\eta$ plane, i.e., the regions
for which the eigenvalues of the $\Omega$ matrix are positive.
The boundaries of the regions are depicted by thick lines.
We note that the regions adjoin each other without overlaps or gaps.
This fact is exact for the ray $\eps=2\eta>0$, the boundary between
the regions of stability for the points FPII and FPIV [at the same
time, this ray is the region of stability for the point FPV].
On the contrary, the boundary $\eps=\eta$, $\eta>0$ for the point FPII
and $\eps=0$, $\eps>\eta$ for FPIV are approximate, so that the
gaps or overlaps can appear in the two-loop approximation.
The regions denoted as FPIV$a$ and FPIV{\it b} with the boundary
$\eps=2$ both correspond to the same fixed point FPIV; the part
FPIV{\it b} represents the region in which the velocity field has
negative critical dimension; see Sec. \ref{sec:summ}.

Surprisingly, Fig.~I has some resemblance with the phase diagrams
presented in Refs. \cite{AvelMaj,Glimm}, despite the essential
difference between the models (in those papers, a strongly
anisotropic velocity field has been studied). Indeed, the boundaries
between the diffusive-type behavior (``homogenization regime'' in
terminology of \cite{AvelMaj}) and convective-type regimes
(``superdiffusive  behavior'') in the two models coincide
(however, in our case they are not exact and will be affected
by the $O(\eps)$ corrections). Furthermore, the Kolmogorov point
($\eps=8/3$, $\eta=4/3$) in our case and in \cite{AvelMaj} lies
on a boundary between two nontrivial regimes. We also note that
the boundary $2\eta=\eps$ between the rapid-change and frozen regimes
was anticipated on phenomenological grounds in Ref. \cite{Falk3},
see also \cite{Eyink}; their arguments can be linked directly
to the RG analysis (see below).

It is clear from the definition of the parameters $g'$, $g''$
that the critical regime governed by the point FPII corresponds to
the rapid-change limit (\ref{RC1}) of our model, while the point
FPIV corresponds to the limit of the frozen velocity field; see Eq.
(\ref{RC2}). This shows that in the latter case, the temporal
fluctuations of the velocity field are asymptotically irrelevant
in determining the inertial-range behavior of the scalar, which
is then completely determined by the equal-time velocity statistics.
In the former case, spatial and temporal fluctuations are both relevant,
but the effective correlation time of the scalar field becomes so
large under renormalization that the correlation time of the velocity
can be completely neglected. The inertial-range behavior of the scalar
is determined solely by the $\omega=0$ mode of the velocity field;
this is the case of the rapid-change model.

We then expect that all the critical dimensions at the point
FPII [FPIV] depend on the only exponent $\zeta\equiv\eps-\eta$
[$\eps$] that survives in the limit in question, and coincide
with the corresponding dimensions obtained directly for
the models (\ref{RC1}) [(\ref{RC2})]. This is indeed the case;
see Eqs. (\ref{DeltaOmega}) and  (\ref{Dnp}) below.

In the regimes governed by the trivial fixed points FPI and FPIII,
the contribution of the convection dies out in the
IR asymptotic region; the IR behavior has purely diffusive
character, while the convection can be treated within ordinary
perturbation theory. The existence of the two fixed points, the frozen
and the rapid-change ones, implies that for $\eta<0$ transport by
{\it small} wavenumbers $k\to0$ is governed by equal-time (spatial)
velocity statistics, while for $\eta>0$ transport by small wavenumbers
is determined by the $\omega=0$ mode, i.e., the time decorrelated component
of the velocity field. If there were IR singularities in
the scalar correlations, they would be determined by the contributions
of small momenta, and these two regimes would be really different.
However, in the regions of stability of the trivial fixed points
there are no such singularities (see the discussion in Sec. \ref{sec:FT}).
Moreover, in these regimes all momenta $k$ contribute to the
long-term, large-scale transport properties of the scalar field
(we recall that for $\eta>0$, $\eps<0$ and $\eta<0$, $\eps<\eta$, the
actual UV cutoff $\Lambda$ has to be introduced, see Sec. \ref{sec:FT},
and the main contribution to the perturbative diagrams then comes from
the momenta of order $k\sim\Lambda$). The RG is not suitable for
studying such ``$\Lambda$ divergent,''  analytic in momenta and frequecies,
quantities. Therefore, the splitting of the homogenization regime
into the rapid-change and frozen parts is not meaningless, but not
practically useful. Probably for this reason it was not mentioned in
Refs. \cite{AvelMaj,AvelMaj2,Glimm}. In what follows, we shall focus
our attention on the nontrivial (anomalous) regimes.

The solution of the RG equations in conformity with the stochastic
hydrodynamics is discussed in Refs. \cite{JETP,UFN,turbo} in detail;
see also \cite{RG,RG1} for the case of the rapid-change models.
Below we restrict ourselves with the only information we need.

Any solution of the RG equation (\ref{RG1}) can be represented in
terms of invariant variables $\bar g(k)$, $\bar u(k)$, and $\bar \nu(k)$,
i.e., the first integrals normalized at $k=\mu$ to $g$, $u$, and $\nu$,
respectively
(we recall that $\mu $ is the renormalization mass in the MS scheme).
The relation between the bare and invariant charges has the form
\begin{equation}
g_{0}=k^{\eps+\eta}\, \bar g\, Z_{g}(\bar g,\bar u),\quad
u_{0}=k^{\eta}\, \bar u\, Z_{u}(\bar g,\bar u), \quad
\nu_0=\bar\nu Z_{\nu}(\bar g,\bar u),
\label{exo1}
\end{equation}
see, e.g., \cite{UFN,turbo,AV}. Equation (\ref{exo1}) determines
implicitly the invariant variables as functions of the bare parameters;
it is valid because both sides of it satisfy the RG equation, and
because Eq. (\ref{exo1}) at $k=\mu$ coincides with (\ref{mult}) owing
to the normalization of the invariant variables.

Correlation time of the velocity field at the wavenumber $k$ is determined
by the relation $t_{v}^{-1}(k) = R(k) = u_{0}\nu_0 k^{2-\eta}$,
see Eqs. (\ref{Fin1}), (\ref{Fin}). Correlation time of the
free scalar field is given by $t_{\theta}^{-1}(k) =  \nu_0 k^{2}$,
in the presence of advection it is replaced by the exact expression
$t_{\theta}^{-1}(k) =\bar \nu(k) k^{2}$. The relations (\ref{svaz})
and (\ref{exo1}) allow the bare parameters and renormalization constants
to be eliminated from the ratio $t_{\theta}(k) /t_{v}(k) $; this gives
\begin{equation}
t_{\theta}(k)/t_{v}(k)=\bar u(k)\propto\const\, k^{-\eta+\gamma^{*}_{\nu}}.
\label{times}
\end{equation}
The last relation in Eq. (\ref{times})  holds for $k\to0$. It follows
from the RG equation ${\cal D}_{k}\bar u=\beta_{u}(\bar g,\bar u)$,
which reduces to ${\cal D}_{k}\bar u= \bar u [-\eta+\gamma^{*}_{\nu}]$
near a fixed point; see Eq. (\ref{beta2}).  Equation
(\ref{times}) discloses the precise physical meaning of the invariant
variable $\bar u$: the ratio of the velocity and scalar correlation
times at the wavenumber $k$. Now we can complete the above discussion of
the scaling regimes and relate it to the  phenomenological arguments
given in Refs. \cite{Falk3,Eyink}.
From (\ref{times}) it follows that for the fixed points FPI and FPII
the velocity correlation time $t_{v}(k)$ becomes very small in comparison
to $t_{\theta}(k)$ for $k\to0$ and can be disregarded; we arrive at the
time-decorrelated velocity field. For FPIII and FPIV, the opposite
inequality, $t_{v}(k)>>t_{\theta}(k)$, holds for small momenta,
the temporal fluctuations of  the velocity are ``frozen in,''
and its correlation time can be replaced with $t_{v}(k)=\infty$.
[Using the representation (\ref{times}) and the exact expressions
for $\gamma_{\nu}^{*}$ in Eqs. (\ref{FPI}), (\ref{FPII}), (\ref{FPIII})
and (\ref{FPIV}), one can easily check that $\bar u\to\infty$ for FPI
and FPII and $\bar u\to0$ for FPIII and FPIV, in agreement with the
analysis of the $\Omega$ matrix.] However, these strong inequalities
for the correlation times hold only asymptotically for $k\to0$, and
therefore the exact correlator (\ref{Fin1}) can be replaced with
its limits (\ref{RC1}) or (\ref{RC2}) only in calculation of a
quantity dominated by small $k$ modes of the velocity field.
Finally, for the point FPV one has $\gamma_{\nu}^{*}=\eta$ and the
ratio (\ref{times}) remains finite for $k\to0$; this is the case of
the local turnover exponent, studied in \cite{Falk3}.

Let $F$ be some multiplicatively renormalized quantity (a parameter,
a field or composite operator), i.e., $F=Z_{F}F_{\rm ren}$ with
certain renormalization constant $Z_{F}$. Then its critical dimension
is given by the expression
\begin{equation}
\Delta[F]\equiv\Delta_{F} = d_{F}^{k}+ \Delta_{\omega}
d_{F}^{\omega}+\gamma_{F}^{*},
\label{32B}
\end{equation}
see, e.g., \cite{JETP,UFN,turbo,Pismak}.
Here $d_{F}^{k}$ and $d_{F}^{\omega}$ are the corresponding canonical
dimensions, $\gamma_{F}^{*}$ is the value of the anomalous dimension
$\gamma_{F}(g)\equiv \widetilde{\cal D}_\mu \ln Z_{F}$  at the fixed
point in question, and $\Delta_{\omega}=2-\gamma^{*}_{\nu}$ is the
critical dimension of frequency. For the nontrivial fixed points
we obtain
\begin{equation}
\Delta_{\omega}= 2-
\cases{ \zeta  &  for FPII, \cr \eps/2  &  for FPIV, \cr
\eta=\eps/2 &   for FPV \cr }
\label{DeltaOmega}
\end{equation}
(we recall that $\zeta\equiv \eps-\eta$, see (\ref{RC1})).
The critical dimensions of the fields $\Phi$ in
our model are also found exactly:
\begin{equation}
\Delta_{\bf v}=1-\gamma^{*}_{\nu},\qquad \Delta_{\theta} = -1
\qquad  \Delta_{\theta'} = d+1,
\label{DimFi}
\end{equation}
and for the IR scale we have $\Delta_{m}=1$ [we recall that all these
quantities in the model (\ref{action}) are not renormalized and
therefore their anomalous dimensions vanish identically,
$\gamma_{\Phi,m}\equiv 0$]. It is also not too difficult to show that
the composite operator $\theta^{n}$ in the model (\ref{action})
is not renormalized and therefore its critical dimension is given
simply by the relation $\Delta[\theta^{n}]= n \Delta[\theta]$; cf.
\cite{RG} for the rapid-change case.

We note that the canonical dimensions of the fields $\theta$,
$\theta'$ in our model (see Table I) differ from their counterparts
in the isotropic rapid-change model (see Table I in Ref. \cite{RG}).
As a result, the critical dimensions (\ref{DeltaOmega}) and
(\ref{DimFi}) at the point FPIV differ from their analogs for
the rapid-change model, in spite of the fact that the anomalous
dimensions are identical. In principle, the canonical dimensions
in two models can be made equal by an appropriate rescaling of
the scalar fields; we shall not dwell on this point here.

Let $G(r)=\langle F_{1}(x)F_{2}(x')\rangle$ be an equal-time
two-point quantity, for example, the pair correlation function
of the primary fields $\Phi$ or some multiplicatively renormalizable
composite operators. The existence of a nontrivial IR stable
fixed point implies that in the IR asymptotic region $\Lambda r>>1$
and any fixed $mr$ the function $G(r)$ takes on the form
\begin{equation}
G(r) \simeq  \nu_{0}^{d_{G}^{\omega}}\, \Lambda^{d_{G}}
(\Lambda r)^{-\Delta_{G}}\, \xi(mr),
\label{RGR}
\end{equation}
with the values of the critical dimensions that correspond to the
fixed point in question and certain scaling function $\xi$ whose
explicit form is not determined by the RG equation itself.
The canonical dimensions $d_{G}^{\omega}$, $d_{G}$ and the
critical dimension $\Delta_{G}$ ot the function $G(r)$ are equal
to the sums of the corresponding dimensions of the quantities
$F_{i}$.

 \section{Critical dimensions of the composite operators
 $\partial\theta\cdots\partial\theta$} \label {sec:Operators}

In the following, an important role will be played by the composite
operators of the form
\begin{equation}
F[n,p]\equiv \partial_{i_{1}}\theta\cdots\partial_{i_{p}}\theta\,
(\partial_{i}\theta\partial_{i}\theta)^{l},
\label{Fnp}
\end{equation}
where $p$ is the number of the free vector indices and $n=p+2l$
is the total number of the fields $\theta$ entering into the operator;
the vector indices of the symbol $F[n,p]$ are omitted.

Coincidence of the field arguments in Green functions containing a composite
operator $F$ gives rise to additional UV divergences. They are removed
by a special renormalization procedure, described in detail, e.g., in
\cite{Zinn,book3,Collins}. The discussion of the renormalization of
composite operators in turbulence models can be found in \cite{UFN,turbo};
see also Ref. \cite{RG} for the case of Kraichnan's model.
Owing to the renormalization, the critical dimension $\Delta[F]$
associated with certain operator $F$ is not in general equal to the
simple sum of critical dimensions of the fields and derivatives entering
into $F$. As a rule, the renormalization of composite operators involves
mixing, i.e., an UV finite renormalized operator is a linear combination
of unrenormalized operators, and vice versa.

The analysis of UV divergences is related to the analysis of the
corresponding canonical dimensions, cf. Sec. \ref{sec:RG}.
It shows that the operators
$F[n,p]$ mix only with each other in renormalization, with the
multiplicative matrix renormalization of the form
(dropping the vector indices everywhere)
\begin{equation}
F[n,p] = Z_{[n,p]\,[n',p']} F_{\rm ren}[n',p']
\label{Matrix}
\end{equation}
Here $F_{\rm ren}$ is the renormalized analog of the operator $F$
and $Z$ is the matrix of renormalization constants. The
corresponding matrix of anomalous dimensions is defined as
\begin{equation}
\gamma_{[n,p]\,[n',p']} = Z_{[n,p]\,[n'',p'']}^{-1}  \Dm
Z_{[n'',p'']\,[n',p']}.
\label{Ma2}
\end{equation}

A simple analysis of the diagrams shows that the matrix element
$Z_{[n,p]\,[n',p']}$ is proportional to $h^{n-n'}$, so that the
elements with $n<n'$ vanish (the parameter $h\equiv|\h|$ appears only
in the numerators of the diagrams; see Sec. \ref{sec:RG}). The
elements with $n=n'$ are independent of $h$ and therefore they can
be calculated directly in the isotropic
model with $\h=0$. The block $Z_{[n,p]\,[n,p']}$ can be then
diagonalized by the changing to irreducible operators (scalars,
vectors, and traceless tensors); but for our purposes it is sufficient
to note that the elements $Z_{[n,p]\,[n,p']}$ vanish for $p<p'$
[the irreducible tensor of the rank $p$ consists of the monomials
with $p'\le p$ only, and therefore only these monomials can admix to
the monomial of the rank $p$ in renormalization]. Therefore, the
renormalization matrix in Eq. (\ref{Matrix}) is triangular, and so is
the matrix (\ref{Ma2}). The isotropy is violated for $\h\ne0$, so
that the irreducible tensors with different numbers of the fields
$\theta$ can mix with each other even though their ranks are also
different. In particular, the vector $\partial_{i}\theta$
admixes to the irreducible tensor
$\partial_{i}\theta\partial_{j}\theta-\delta_{ij}(\partial_{s}\theta
\partial_{s}\theta)/d$ in the form of the traceless combination
$2\delta_{ij} (h_{s}\partial_{s}\theta)/d-h_{i}\partial_{j}\theta-
h_{j}\partial_{i}\theta$. In the following, we shall not be interested
in the precise form of the basis operators, i.e., those having
definite anomalous dimensions; we shall rather be interested in the
anomalous dimensions themselves. The latter are given by the
eigenvalues $\gamma[n,p]$
of the matrix (\ref{Ma2}), and in our case they are completely
determined by the diagonal elements of the renormalization matrix
(\ref{Matrix}): $\gamma[n,p]= \Dm \ln Z_{[n,p]\,[n,p]}$.

Now let us turn to the one-loop calculation of the constant
(\ref{Znp}) in the MS scheme.
Let $\Gamma(x;\theta)$ be the generating functional of the
1-irreducible Green functions with one composite operator $F[n,p]$
from Eq. (\ref{Fnp}) and any number of fields $\theta$. Here $x\equiv
(t,{\bf x})$ is the argument of the operator and $\theta(x)$ is
the functional argument, the ``classical analog'' of the random
field $\theta$. We are interested in the $\theta^{n}$ term of the
expansion of $\Gamma(x;\theta)$ in $\theta(x)$, which we denote
$\Gamma_{n}(x;\theta)$; it has the form
\begin{equation}
\Gamma_{n}(x;\theta) = \frac{1}{n!} \int dx_{1} \cdots \int dx_{n}
\, \theta(x_{1})\cdots\theta(x_{n})\,
\langle F[n,p](x) \theta(x_{1})\cdots\theta(x_{n})\rangle_{\rm 1-ir}.
\label{Gamma1}
\end{equation}
In the one-loop approximation the function (\ref{Gamma1}) is
represented diagramatically in the following manner:
\begin{equation}
\Gamma_{n}= F[n,p] +\frac{1}{2} \put(-20.00,-50.00){\makebox{\dA}}
\hskip1.4cm .
\label{Gamma2}
\end{equation}
The first term is the ``tree'' approximation, and the black circle with
two attached lines in the diagram denotes the variational derivative
$\delta^{2} F[n,p] / {\delta\theta\delta\theta}$. In the momentum
representation it has the form
\begin{eqnarray}
T ({\bf k},{\bf q})\equiv
\frac{\delta^{2} F[n,p]}{\delta\theta({\bf k})\delta\theta({\bf q})} =
- p\,(p-1)\, k_{i_{1}}q_{i_{2}}\,
(\partial_{i_{3}}\theta\cdots\partial_{i_{p}}\theta)\,
(\partial_{i}\theta\partial_{i}\theta)^{l} \nonumber \\
- 4pl\, k_{i_{1}}q_{s}\,(\partial_{i_{2}}\theta\cdots
\partial_{i_{p}}\theta)\, \partial_{s}\theta \,
(\partial_{i}\theta\partial_{i}\theta)^{l-1} \nonumber \\
- 2l ({\bf k}\cdot{\bf q})\,
(\partial_{i_{1}}\theta\cdots\partial_{i_{p}}\theta)\,
(\partial_{i}\theta\partial_{i}\theta)^{l-1} \nonumber \\
- 4l(l-1)\,  k_{j}q_{s} (\partial_{j}\theta\partial_{s}\theta)\,
(\partial_{i_{1}}\theta\cdots\partial_{i_{p}}\theta)\,
(\partial_{i}\theta\partial_{i}\theta)^{l-2}.
\label{variat}
\end{eqnarray}
Strictly speaking, we had to symmetrize the right-hand side of Eq.
(\ref{variat}) with respect to the indices $i_{1}\cdots i_{p}$ and
the momenta ${\bf k}$, ${\bf q}$. However, the symmetry is restored
automatically after the vertex $T ({\bf k},{\bf q})$ has been
inserted into the diagram, which is why only one term of each type
is displayed in Eq. (\ref{variat}) and the required symmetry
coefficients are introduced.

The vertex (\ref{variat}) contains $(n-2)$ factors of $\partial\theta$.
Two remaining ``tails'' $\theta$  are attached to the vertices
$\theta'({\bf v}\partt)\theta$ of the diagram (\ref{Gamma2}).
It follows from the explicit form of the vertices that these two
fields $\theta$ are isolated from the diagram in the form of the
overall factor $\partial\theta\partial\theta$; cf. Sec. \ref{sec:RG}.
In other words, two external momenta, corresponding to these fields
$\theta$, occur as an overall factor in the diagram, and the
UV divergence of the latter is logarithmic rather than quadratic;
cf. the expression (\ref{novaja2}), (\ref{Dyson4}). Therefore, we
can set all the external momenta and the ``mass'' $m$ equal to zero
in the integrand; the IR regularization is provided by the cut-off
of the integral at $q\simeq m$.
Then the UV divergent part of the one-loop diagram (\ref{Gamma2})
can be written in the form
\begin{equation}
\partial_{p}\theta\partial_{l}\theta
\int \frac{d\omega}{2\pi} \int
\frac{d{\bf q}}{(2\pi)^{d}} \, T ({\bf q},-{\bf q})
\frac{P_{pl}({\bf q})\,D_{v}(\omega,q)} {\omega^{2}+\nu^2 q^{4}} .
\label{diagr01}
\end{equation}
The expression (\ref{diagr01}) is a linear
combination of the integrals
\begin{equation}
T_{ij,pl}= \int \frac{d\omega}{2\pi} \int
\frac{d{\bf q}}{(2\pi)^{d}} \, \frac{q_{i}q_{j}\,
P_{pl}({\bf q})\,D_{v}(\omega,q)} {\omega^{2}+\nu^2 q^{4}}.
\label{diagr1}
\end{equation}
We perform the integration over $\omega$ and make use of the
isotropy, namely,
\[ \int d{\bf q}\, f(q)\,q_{i}q_{j}\, P_{pl}({\bf q}) =
\frac{(d+1)\delta_{pl}\delta_{ij}-\delta_{pi}\delta_{lj} -
\delta_{pj}\delta_{li}}{d(d+2)} \int d{\bf q}\, f(q)\, q^{2}. \]
This gives
\begin{equation}
T_{ij,pl}=
\frac{(d+1)\delta_{pl}\delta_{ij}-\delta_{pi}\delta_{lj} -
\delta_{pj}\delta_{li}}{2ud(d+2)}\,
g\mu^{\eps} \,J
\label{diagr2}
\end{equation}
with the integral $J$ from Eq. (\ref{J}).

Substituting Eqs. (\ref{variat}) and (\ref{diagr2}) into Eq.
(\ref{diagr01}) gives the desired expression for the divergent
part of the diagram (\ref{Gamma2}). In this expression we have
to take into account all the terms proportional to the operator
$F[n,p]$ and neglect all the other terms, namely, the terms
containing the factors of $\delta_{i_{1}i_{2}}$ etc. The latter
determine non-diagonal elements of the matrix (\ref{Matrix}),
which we are not interested in here. Finally we obtain
\begin{equation}
\Gamma_{n}\simeq F[n,p] \left[1 - \frac {g\mu^{\eps} \,J\,Q[n,p]}
{4ud(d+2)}  \right] + \cdots,
\label{diagr3}
\end{equation}
where we have written
\begin{eqnarray}
Q _{np}& \equiv & 2n\,(n-1) - (d+1)\, (n-p)\, (d+n+p-2) =
\nonumber \\  &=& 2p\,(p-1) - (d-1)\, (n-p)\, (d+n+p) .
\label{Qnp}
\end{eqnarray}
The dots in Eq. (\ref{diagr3}) stand for the $O(g^{2})$ terms and
the structures different from $F[n,p]$, $\simeq$ denotes the
equality up to the UV finite parts; we also recall that $n=p+2l$.

The constant $Z_{[n,p],[n,p]}$ is found from the requirement
that the renormalized analog $\Gamma_{n}^{\rm ren}\equiv
Z_{[n,p],[n,p]}^{-1}\Gamma_{n}$ of the function (\ref{diagr3})
be UV finite (mind the minus sign in the exponent); along with
the representation (\ref{J1}) for the integral $J$ and the MS
scheme this gives the following result:
\begin{equation}
Z_{[n,p]\,[n,p]}=1- \frac{ag}{u}\, \frac{Q[n,p]}{2(d-1)(d+2)}
\sum^{\infty}_{s=0} \frac{(-u)^{s}}
{\eps+s\eta}\, ,
\label{Znp}
\end{equation}
with the polynomial $Q[n,p]$ from Eq. (\ref{Qnp})
and the constant $a$ is given in Eq. (\ref{a}).

For the anomalous dimension (\ref{Ma2}) we then obtain:
\begin{equation}
\gamma[n,p]=  \frac{ag\,Q[n,p]}{2u(u+1)(d-1)(d+2)}\, ;
\label{Q}
\end{equation}
cf. Sec. \ref{sec:RG} for the dimension $\gamma_{\nu}$.
The critical dimension associated with the operator $F[n,p]$
has the form $\Delta[n,p]= \gamma^{*}[n,p]$; see Eq. (\ref{32B})
and Table I ($\gamma^{*}$ denotes the value of $\gamma$ at the fixed
point in question). For the nontrivial fixed points discussed in
Sec. \ref{sec:Fixed} we then obtain
\begin{equation}
\Delta[n,p] = \frac{Q[n,p]}{2(d-1)(d+2)}\times \cases{
\zeta\equiv\eps-\eta  &  for FPII, \cr
\eps/2  &  for FPIV, \cr
\eta =\eps/2 &   for FPV \cr }
\label{Dnp}
\end{equation}
with the corrections of order $O(\eps^{2})$.

The expression (\ref{Dnp}) illustrates the general fact that the
critical dimensions in the rapid-change and frozen regimes depend
only on the exponents $\zeta$ and $\eps$, respectively.
It turns out that the dimension  $\Delta[n,p]$ at the point FPV
is universal, i.e., it is independent of the free parameter
$u_{*}$, or, equivalently, of the specific choice of a fixed point
on the curve described by Eq. (\ref{FPV}). This is a consequence of
the explicit form of the RG functions in the one-loop approximation
(the same combination $g/u(u+1)$ enters into the beta functions and
the anomalous dimension of the operator $F[n,p]$). We then expect that
the exact dimension $\Delta[n,p]$ at the point FPV is nonuniversal,
and the dependence on $u_{*}$ will appear at the two-loop level.
Another artifact of the one-loop approximation  is the continuity
of the dimension $\Delta[n,p]$ at the crossover line
$\eps=2\eta$ as a function of the exponents $\eps$, $\eta$.

The first-order result (\ref{Dnp}) for the operator $F[2,0]$ (the
local dissipation rate) is in fact exact. The proof is based on
certain Schwinger equation; it is almost identical to the analogous
proof for the Kraichnan model, given in \cite{RG}, and will not
be discussed here.

The above analysis applies also to the case of a nonsolenoidal velocity
field (compressible fluid). The transversal projector in Eq. (\ref{f})
is then replaced with $P_{ij}({\bf k})+\alpha Q_{ij}({\bf k})$,
where $Q_{ij}({\bf k})\equiv k_{i}k_{j}/k^{2}$ is the longitudinal
projector and $\alpha>0$ is an additional arbitrary parameter,
the degree of compressibility. For the rapid-change regime (\ref{RC1}),
the dimension $\Delta[n,p]$ takes on the form
\begin{equation}
\Delta[n,p] = \frac{-\zeta}{(d+2)} \left[ \frac{(n-p)(d+n+p)}{2} +
\frac{p(p-1)(\alpha-1)+\alpha(n-p)(n+p-2)}{(d-1+\alpha)}
 \right] + O(\zeta^{2}),
\label{Dnp2}
\end{equation}
in agreement with the $p=0$ results obtained in Refs. \cite{tracer}
for the `tracer' model and earlier in Ref. \cite{VM} for $d=1$.
In general case (\ref{Fin}),
additional superficial UV divergence emerges in the 1-irreducible
Green function  $\langle\theta'\theta v\rangle_{\rm 1-ir}$,
and the second independent renormalization constant should be
introduced as a coefficient in front of the new counterterm
$\theta'({\bf v}\partt)\theta$. This case requires special analysis
and will be discussed elsewhere\footnote{See: N.~V.~Antonov,
{\it Anomalous scaling of a passive scalar advected by the synthetic
compressible flow,} chao-dyn/9907018.} (in particular, the nontrivial
fixed point becomes infinite for the purely potential frozen
velocity field, cf. \cite{walks1,walks3} for the random walks
in random environment).

 \section{Operator product expansion and the anomalous scaling for
 the structure functions and other correlators} \label {sec:OPE}

The representation (\ref{RGR}) for any scaling function $\xi(mr)$
describes the behavior of the Green function for $\Lambda r>>1$
and any fixed value of $mr$. The inertial-convective range corresponds
to the additional condition that $mr<<1$. The form of the function
$\xi(mr)$ is not determined by the RG equations themselves; in
the theory of critical phenomena, its behavior for $mr\to0$ is
studied using the well-known Wilson operator product expansion (OPE);
see, e.g., \cite{Zinn,book3,Collins}. This technique is also applicable
to the theory of turbulence; see \cite{RG,RG1,LOMI,JETP,UFN,turbo}.

According to the OPE, the equal-time product $F_{1}(x)F_{2}(x')$
of two renormalized operators at
${\bf x}\equiv ({\bf x} + {\bf x'} )/2 = {\const}$ and
${\bf r}\equiv {\bf x} - {\bf x'}\to 0$
has the representation
\begin{equation}
F_{1}(x)F_{2}(x')=\sum_{\alpha}C_{\alpha} ({\bf r})
F_{\alpha}({\bf x,t}) ,
\label{OPE}
\end{equation}
where the functions $C_{\alpha}$  are the Wilson coefficients
regular in $m^{2}$ and  $ F_{\alpha}$ are all possible
renormalized local composite operators allowed by symmetry, with
definite critical dimensions $\Delta_{\alpha}$.
The renormalized correlator $\langle F_{1}(x)F_{2}(x') \rangle$
is obtained by averaging Eq. (\ref{OPE}) with the weight
$\exp S_{ren}$, the quantities  $\langle F_{\alpha}\rangle$
appear on the right-hand side. Their asymptotic behavior
for $m\to0$ is found from the corresponding RG equations and
has the form $\langle F_{\alpha}\rangle \propto  m^{\Delta_{\alpha}}$.
From the operator product expansion (\ref{OPE}) we therefore
find the following expression  for the scaling function
$\xi(mr)$ in the representation (\ref{RGR}) for the correlator
$\langle F_{1}(x)F_{2}(x') \rangle$:
\begin{equation}
\xi(mr)=\sum_{\alpha}A_{\alpha}\,(mr)^{\Delta_{\alpha}},
\label{OR}
\end{equation}
where the coefficients $A_{\alpha}=A_{\alpha}(mr)$ are regular
in $(mr)^{2}$.

In the models of critical phenomena, the leading contribution to the
representations like (\ref{OR}) is related to the simplest operator
$F=1$ with the minimal dimension $\Delta_{\alpha}=0$, while the other
operators determine only the corrections that vanish for $mr\to0$.
The feature characteristic of the turbulence models is the existence
of the so-called ``dangerous'' composite operators with negative
critical dimensions \cite{RG,RG1,LOMI,JETP,UFN,turbo}. Their contributions
to the operator product expansions determine the IR behavior of the
scaling functions and lead to their singular dependence on $m$
for $mr\to0$.

If the spectrum of the dimensions $\Delta_{\alpha}$ for a given
scaling function is bounded from below, the leading term of its
behavior for $mr\to0$ is simply given by the minimal dimension.
This is the case of the rapid-change model (see \cite{RG,RG1}),
and, as we shall see below, of our model (\ref{action}).
[The exception is provided by the non-rapid-changes regimes, if
the values of the exponents $\eps$, $\eta$ are large enough. It
is discussed in the subsequent Section.]

Consider for definiteness the equal-time structure functions
of the scalar field:
\begin{equation}
S_{n}(r)\equiv\langle[\theta(t,{\bf x})-\theta(t,{\bf x'})]^{n}\rangle,
\quad  r\equiv|{\bf x}-{\bf x'}| .
\label{struc}
\end{equation}
For the functions (\ref{struc}), the representation of the form
(\ref{RGR}) is valid with the dimensions $d_{G}^{\omega}=0$ and
$d_{G}=\Delta_{G}=n \Delta_{\theta}=-n$. In general, the operators
entering into the operator product expansions are
those which appear in the corresponding Taylor expansions, and also
all possible operators that admix to them in renormalization.
The leading term of the Taylor expansion for the function (\ref{struc})
is given by the $n$-th rank tensor $F[n,n]$ from Eq. (\ref{Fnp}).
The decomposition of $F[n,n]$ in irreducible tensors gives rise to
the dimensions $\Delta[n,p]$ with all possible values of $p$;
the admixture of junior operators gives rise to all the dimensions
$\Delta[k,p]$  with $k<n$. Therefore, the asymptotic expression for
the structure function has the form
\begin{equation}
S_{n}(r) \simeq (hr)^n\,  \sum_{k=0}^{n} \sum_{p=p_{k}}^{k}
\left[C_{kp}\, (mr)^{\Delta[k,p]}+\cdots\right].
\label{struc2}
\end{equation}
Here and below $p_{k}$ denotes the minimal possible value of $p$ for
given $k$, i.e., $p_{k}=0$ for $k$ even and $p_{k}=1$ for $k$ odd;
$C_{kp}$ are some numerical coefficients dependent on $\eps$, $\eta$,
$d$, and on the angle between the vectors ${\bf h}$ and ${\bf r}$.

Some remarks are in order.

The dots in Eq. (\ref{struc2}) stand for the contributions of the order
$(mr)^{2+O(\eps)}$ and higher, which arise from the senior operators,
for example, $\partial^{2}\theta\partial^{2}\theta$ or ${\bf v}^2$.

In the original Kraichnan model, only scalar operators give
contributions to the representations like (\ref{struc2}),
because the mean values $\langle F_{\alpha}\rangle$ of all
the other irreducible tensors vanish owing to the isotropy;
see \cite{RG,RG1}. In the model (\ref{action}), the traceless
irreducible tensors acquire nonzero mean values, and their
dimensions appear on the right-hand side of Eq. (\ref{struc2}).
In particular, the mean value of the operator
$\partial_{i}\theta\partial_{j} \theta-\delta_{ij}
(\partial_{s}\theta \partial_{s}\theta)/d$ is proportional to the
traceless tensor of the form $\delta_{ij}(h_{s}h_{s})/d-h_{i}h_{j}$,
its tensor indices are contracted with the indices of the
corresponding coefficient $C_{\alpha}$ in Eq. (\ref{OPE}).

The operators $F[k,p]$ with $k>n$ (whose contributions would be
more important) do not appear in Eq. (\ref{struc2}), because
they do not appear in the Taylor expansion of the function $S_{n}$
and do not admix in renormalization to the terms of the Taylor
expansion.

The leading term of the expression (\ref{struc2}) for $mr\to0$
is obviously given by the contribution with the minimal possible
dimension. The straightforward analysis of the explicit one-loop
expression (\ref{Dnp}) shows that for fixed $n$, any $d\ge2$, and
any nontrivial fixed point, the dimension $\Delta[n,p]$
decreases monotonically with $p$ and reaches
its minimum for the minimal possible value of $p=p_{n}$, i.e.,
$p=0$ if $n$ is even and $p=1$ if $n$ is odd. Furthermore, this
minimal value $\Delta[n,p_{n}]$ decreases monotonically as $n$
increases, i.e.,
\[ \Delta[2k,0]>\Delta[2k+1,1]>\Delta[2k+2,0]. \]
[A similar hierarchy has been established recently in Ref. \cite{hi}
for the magnetic field advected passively by the rapid-change velocity
in the presence of large-scale anisotropy.]
Therefore, the desired leading term for the even (odd) structure
function $S_n$ is determined by the scalar (vector) composite
operator consisting of $n$ factors $\partial\theta$
and has the form
\begin{equation}
S_{n}(r) \propto (hr)^n\, (mr)^{\Delta[n,p_{n}]}
\label{struc3}
\end{equation}
with the dimension $\Delta[n,p]$ given in Eq. (\ref{Dnp}).

For the rapid-change fixed point and even values of $n$, the
total power of $r$ in Eq. (\ref{struc3}) coincides with the
exponent in the original isotropic Kraichnan model, calculated
to the order $O(\zeta)$ in \cite{BGK} and $O(1/d)$ in \cite{Falk2}
within the zero-mode approach, and to the order $O(\zeta^{2})$
in \cite{RG} within the framework of the RG. We also note that
the anomalous dimensions associated with the operators $F[2k,2]$
were calculated in \cite{RG} to the order $O(\zeta^{2})$; the
exact dimension of the operator $F[2,2]$ was found in \cite{Falk1}.
[It should be noted that the {\it decomposition} of the total
exponent in Eq. (\ref{struc3}) into the critical dimension of the
composite operator and the critical dimension of the structure
function itself differs from the analogous decomposition for
Kraichnan's model, as a result of the difference in canonical
dimensions; see Sec. \ref{sec:RG}].

The result (\ref{struc3}) for the third-order structure function
in the rapid-change model coincides with the $O(\zeta)$ result
obtained in \cite{Pumir2} within the zero-mode technique; see also
the earlier paper \cite{Pumir} for the three-dimensional result. We
note that the exponents $-7\zeta/5$ and $3\zeta/5$ from \cite{Pumir}
should be identified with the anomalous dimensions $\Delta[3,1]$ and
$\Delta[3,3]$, respectively. The result (\ref{struc3}) for $n=3$ is
also in agreement with the $O(1/d)$ result obtained in \cite{Gut},
with the identification $\gamma+1-\Delta=3+\Delta[3,1]$.

For the case of the frozen velocity field (FPIV), the first-order
results for the even structure functions were presented in \cite{RG}.
We also note that they satisfy the exact inequalities obtained for the
time-independent case in \cite{Eyink}.

The analysis given above is extended directly to the case of other
correlation functions. For example, the analog of the expression
(\ref{struc3}) for the equal-time pair correlation function of the
operators (\ref{Fnp}) has the form
\begin{equation}
\langle F[n,p]\, F[n',p']\rangle \simeq  h^{n+n'}\,
(\Lambda r)^{-\Delta[n,p_{n}]-\Delta[n',p_{n'}]}
(mr)^{\Delta[n+n',\, p_{n+n'}]} \, .
\label{struc4}
\end{equation}
Some special cases of the relation (\ref{struc4}) for the
rapid-change model were obtained earlier in Refs.
\cite{Falk1,Falk2,GK,BGK,RG}.

Another interesting example is the equal-time pair correlator
$\langle\theta^{n}(t,{\bf x})\theta^{k}(t,{\bf x'})\rangle$.
Substituting the relations $d^{\omega}_{G}=0$ and
$d_{G}=\Delta_{G}=-(n+k)$ into the general expression (\ref{RGR}) gives
$\langle\theta^{n}\theta^{k}\rangle=r^{n+k}\xi(mr)$, and the small
$mr$ behavior of the scaling function $\xi(mr)$ is found from Eq.
(\ref{OR}) (here and below, we do not display the obvious dependence on
$h$). In contrast to the previous example, composite operators
in the expansion (\ref{OPE}) can involve the field $\theta$
{\it without derivatives}. The leading term in Eq. (\ref{OR}) is then
given simply by the operator $\theta^{n+k}$ with $\Delta_{F}=-(n+k)$,
while the first correction is related to the monomial
$(\partial_{i}\theta\partial_{i}\theta)\theta^{n+k-2}$
whose critical dimension is easily found to be $\Delta_{F}=
-(n+k)+\Delta_{\omega}$ with $\Delta_{\omega}$ from Eq.
(\ref{DeltaOmega}). Therefore, in the inertial range our correlator
has the form $\langle\theta^{n}\theta^{k}\rangle\simeq c_{1}
m^{-(n+k)}-c_{2} m^{-(n+k)}(mr)^{\Delta_{\omega}}+\dots$,
a large constant minus a powerlike correction (the signs of the
constants $c_{i}$ are explained by the fact that the correlator is
positive and decreases as $r$ grows). In the structure functions
(\ref{struc}) all the contributions related to operators containing
fields without derivatives cancel out to give the behavior
(\ref{struc2}), determined by the operators constructed only of
field derivatives.

Finally, we note that the hierarchy of critical dimensions $\Delta[n,p]$,
established in Sec.\ref{sec:Operators}, persists also for the
nonsolenoidal velocity field, see (\ref{Dnp2}). Therefore, the
asymptotic expressions like (\ref{struc2}), (\ref{struc3}) and
(\ref{struc4}) remain valid for the compressible case (a `tracer' in
terminology of \cite{tracer}) with the exponents $\Delta[n,p]$
given in (\ref{Dnp2}).

 \section{Summation of the dangerous contributions
 from the powers of the velocity field} \label{sec:summ}

A new interesting problem emerges as the parameters $\eps$ and
$\eta$ increase and the velocity field becomes dangerous; see Eq.
(\ref{DimFi}). Owing to the Gaussianity, all its powers also
become dangerous with the dimensions
$\Delta[v_{i_1}\cdots v_{i_n}]=n\Delta_{\bf v}$.
The analysis of the diagrams shows that in the rapid-change
regime, these operators do not contribute to the operator product
expansions of the equal-time correlators like (\ref{struc}) or
(\ref{struc4}), but all the contributions of the scalars $({\bf
v}^2)^n$ do appear in those OPE for the non-rapid-change regimes.
The spectrum of their dimensions is unbounded from below, and in
order to find the small $mr$ behavior we have to sum up all their
contributions $\propto (mr)^{n\Delta[{\bf v}]}$ in the representation
(\ref{OR}). We have employed the infrared perturbation theory in the
form developed in \cite{LOMI,JETP} to perform the required
summation for the structure function $S_2$ in the frozen regime and
within the one-loop approximation for the Wilson coefficients.
In this case, the velocity becomes dangerous for $\eps>2$ (region
FPIV{\it b} on Fig.~I).

The function $S_2$ is represented in the form
\begin{equation}
S_{2} = \int \D\Phi\, [\theta(t,{\bf x})-\theta(t,{\bf x'})]^{2}
\, \exp S(\Phi)
\label{sum1}
\end{equation}
with the action functional from Eq. (\ref{action}).
Following \cite{LOMI,JETP}, we split the velocity field
in Eq. (\ref{sum1}) into two
components, ${\bf v}(x) = {\bf v}_{<} (x) +{\bf v}_{>} (x)$,
referring to the ``soft'' component,  ${\bf v}_{<}$, all the
Fourier modes with $k\le k_{*}$, and to to the ``hard'' component,
${\bf v}_{>}$, all the remaining modes with $k> k_{*}$. Here
$k_{*}$ is a fixed arbitrary separating scale, which will not
enter into the final expressions. Since we are interested in only
the contributions of the operators $({\bf v}^2)^n$ into the OPE,
we can neglect the spacetime inhomogeneity of the soft field. It
then becomes a random variable (rather than a random field) with
the statistics determined by the relation
\begin{equation}
\langle{\bf v}_{<}\cdots{\bf v}_{<} \rangle\equiv \langle{\bf v} (x)
\cdots{\bf v}(x)\rangle.
\label{sum2}
\end{equation}
Furthermore, we confine ourselves to the one-loop approximation for
the corresponding Wilson coefficients, so that we can omit the
contribution of the hard field in the vertex
$\theta'({\bf v}\partt)\theta$. Then the integration over the fields
$\theta$, $\theta'$, and ${\bf v}_{>}$ in Eq. (\ref{sum1}) gives:
\begin{equation}
S_{2} =2 \int \frac{d{\bf k}}{(2\pi)^{d}}
\int \frac{d\omega}{(2\pi)} \biggl[ 1-\exp({\rm i}{\bf k}\cdot{\bf r})
\biggr] \frac{P({\bf k}) D_{v}(\omega,{\bf k})}
{(\omega-{\bf v}_{<}\cdot{\bf k})^{2}+\nu_0^{2} k^{4}}\, ,
\label{sum3}
\end{equation}
where $P({\bf k})\equiv h_{i}h_{j}P_{ij}({\bf k})$, the correlator
$D_{v}$ is given by Eq. (\ref{Fin}), and the averaging over
${\bf v}_{<}$ with the statistics (\ref{sum2}) is to be performed.
The mean values (\ref{sum2}) in Eq. (\ref{sum3}) correspond to
the contributions from $\langle{\bf v}(x)\cdots{\bf v}(x)\rangle$
in the in the representation (\ref{OR}) for $S_{2}$.
For the rapid-change model, the correlator $D_{v}$ is independent
of the frequency; see Eq. (\ref{RC1}). It then follows from the
expression (\ref{sum3}) that the dependence on ${\bf v}_{<}$ vanishes
after the integration over $\omega$, which means that the operators
$({\bf v}^2)^n$ give no contribution to the OPE for $S_{2}$. Now let
us turn to the case of the frozen velocity field. We substitute the
correlator (\ref{RC2}) into Eq. (\ref{sum3}) and perform the
integration over $\omega$; this gives:
\begin{equation}
S_{2} = g_{0}''\nu_0^{2}\,\int \frac{d{\bf k}}{(2\pi)^{d}}
\biggl[ 1-\exp({\rm i}{\bf k}\cdot{\bf r}) \biggr]
\frac{P({\bf k})\,(k^{2}+m^{2})^{-d/2+1-\eps/2} }
{({\bf v}_{<}\cdot{\bf k})^{2}+\nu_0^{2} k^{4}}.
\label{sum4}
\end{equation}

A brief digression is required here. We are interested in the
small $mr$ behavior of the scaling function $\xi(mr)$ from Eq.
(\ref{OR}), so that we have to combine the expression (\ref{sum4})
with the representation (\ref{RGR}) for the function $S_{2}$.
In renormalized variables, the latter is written in the form
$S_{2}=(hr)^{2} f(\mu r,g'', u, mr)$, where $f$ is some function of
completely dimensionless arguments. The function $\xi(mr)$ is then
given by the relation $\xi(mr)=f(1, g_{*}'', 0, mr)$ with $g_{*}''$
from Eq. (\ref{FPIV}) (we recall that for the frozen regime,
$u_{*}=0$). The renormalization of Eq. (\ref{sum4}) in our
approximation reduces to the replacement
$g_{0}''\to g''\mu^{\eps}$, $\nu_0\to \nu$, and the changeover
to the scaling function is given by the substitution
$g''\to g_{*}''$, $\mu\to 1/r$. From now on, all these substitutions
are implied. The expansion of the denominator in Eq. (\ref{sum4}) in
$({\bf v}_{<}\cdot{\bf k})^{2}$ gives:
\begin{equation}
\frac{1}{({\bf v}_{<}\cdot{\bf k})^{2}+\nu^{2} k^{4}} =
\frac{1}{\nu^{2} k^{4}}  \sum_{n=0}^{\infty} (-1)^{n}\,
\frac{({\bf v}_{<}\cdot{\bf k})^{2n}}{\nu^{2n} k^{4n}}.
\label{sum5}
\end{equation}
It follows from Eqs. (\ref{RC2}) and (\ref{sum2}) that the correlators
of the soft field have the form
\begin{equation}
\langle (v_{<})_{i_{1}}\cdots (v_{<})_{i_{2n}} \rangle = D\,
[\delta_{i_{1}i_{2}}\delta_{i_{3}i_{4}}\cdots\delta_{i_{2n-1}i_{2n}}+
{\rm all\ possible\ permutations}\,] ,
\label{sum6}
\end{equation}
where
\begin{equation}
D\simeq  r^{\eps} \nu^{2}\,\int \frac{d{\bf k}}{(2\pi)^{d}}\,
(k^{2}+m^{2})^{-d/2+1-\eps/2}
\simeq \nu^{2}\,m^{2} (mr)^{-\eps}.
\label{sum7}
\end{equation}
(here and below $\simeq$ denotes the equality up to a numeric factor).
Strictly speaking, the integral (\ref{sum7}) should be cut off from
above at $k\sim k_{*}$. For $\eps>1$, the cut-off can be removed
without changing the singularity at $m\to0$. The averaging Eq.
(\ref{sum5}) over ${\bf v}_{<}$ gives:
\begin{equation}
\frac{1}{\nu^{2}k^{4}}\sum_{n=0}^{\infty} \frac{(2n)!}{n!}\, (-z)^{n},
\quad z\equiv \frac {D} {2k^{2}\nu^2}\simeq \frac {m^{2}(mr)^{-\eps}}{k^{2}}
\label{sum8}
\end{equation}
[we note that $(2n)!/2^{n}n!=(2n-1)!!$ is the number of terms in Eq.
(\ref{sum6})]. The small $mr$ limit implies $z\to\infty$. The
large $z$ behavior of the series in Eq. (\ref{sum8})
is found from the following integral representation:
\begin{equation}
\sum_{n=0}^{\infty} \frac{(2n)!}{n!} (-z)^{n}= \int^{\infty}_{0}
dt\, \exp(-zt^{2}-t) = \sqrt{\pi/4z} \left[1+O(1/\sqrt{z})
\right].
\label{sum9}
\end{equation}
Substituting Eq. (\ref{sum9}) into Eq. (\ref{sum4}) gives:
\begin{equation}
S_{2}\simeq m^{-2} (mr)^{-\eps/2} \int d {\bf y}
\biggl[ 1-\exp({\rm i}{\bf y}\cdot m{\bf r}) \biggr] P({\bf y})
y^{-3} (1+y^{2})^{-d/2+1-\eps/2},
\label{sum10}
\end{equation}
where we have changed to the dimensionless integration variable
${\bf y}$ defined so that ${\bf k}=m{\bf y}$. Expansion of the
quantity in the square brackets for small $mr$ gives
\begin{equation}
S_{2}\simeq (mr)^{-\eps/2} r_{i}r_{j} \int d {\bf y}
 P({\bf y}) y_{i}y_{j} y^{-3} (1+y^{2})^{-d/2+1-\eps/2}
\label{sum11}
\end{equation}
[the first term of the expansion gives no contribution to Eq.
(\ref{sum11}), owing to the evenness of the rest of the integrand].
The vector indices can be isolated from the integral (\ref{sum11});
this gives rise to the two structures, $\delta_{ij}h^{2}$ and
$h_{i}h_{j}$, with the scalar coefficients proportional to the
integral $\int d {\bf y} y^{-1} (1+y^{2})^{-d/2+1-\eps/2}$.
One can easily check that for $\eps+\eta>1$, it is
both IR and UV convergent, so that the
leading terms of the small $mr$ behavior of the integral
(\ref{sum10}) are indeed obtained simply by the expansion of the
integrand and have the form $h^{2}\,r^{2}\,(mr)^{-\eps/2}$
and $({\bf h}\cdot{\bf r})^{2}(mr)^{-\eps/2}$.

Therefore, it turns out that the contributions of the
operators $({\bf v}^2)^n$ sum up to the powerlike expression $r^2
(mr)^{-\eps/2}$. In other words, the infinite sum of these
dangerous operators gives to the function $S_2$ contribution of
the same form as the single operator $F[2,0]$, and therefore the IR
behavior of $S_2$ is given by the same expression (\ref{struc3})
for all values of the exponent $\eps$ [of course, the
corresponding amplitudes for $\eps>2$ acquire an additional
contribution from the operators $({\bf v}^2)^n$].

The infinite family of the dangerous operators $({\bf v}^2)^n$
also occurs in the RG approach to the stochastic NS
equation; see \cite{LOMI,JETP,UFN,turbo}. In that case, their
summation in the OPE for different-time correlators leads to the
expressions that are analogous to \cite{sweep} and describe
the well-known sweeping effects. The contributions of these
operators vanish when one changes to the equal-time Galilean
invariant functions, for example, the structure functions of the
velocity field. In this connection, it should be
stressed that the appearance of the operators $({\bf v}^2)^n$
in the structure functions of the model (\ref{action}) is not an
artifact of the synthetic velocity statistics. One can check that
in the presence of a mean gradient, the same effect occurs even
though the velocity is generated by the Galilean covariant stochastic
NS equation. On the contrary, if the effective scalar
noise in the diffusion equation is proportional to the delta function
in time (as in original Kraichnan's model), the operators
$({\bf v}^2)^n$ are absent from the equal-time correlators whatever be
the velocity statistics. One can probably consider this effect as an
additional source of the breakdown of the Kolmogorov--Obukhov
theory for the passive scalar advection.

The summation given above can be viewed as a possible model of the
origin of the anomalous scaling in the structure functions of the
velocity field: it was argued in \cite{AV} that the singular
dependence on $mr$ of the equal-time correlators for the stochastic
NS equation can be related to {\it infinite} families
of dangerous operators. The summation can also be performed in a more
formal way, without referring to the infrared perturbation theory,
by using only the operator product expansion for the quantity
$[\theta(t,{\bf x})- \theta(t,{\bf x'})]^{2}$ and not only its average
(\ref{sum1}). It is also possible to take into account
all the composite operators constructed of the velocity and its time
derivatives; see \cite{Kim}.

 \section{``Exotic'' scaling regimes} \label {sec:Exo}

So far, we have considered the standard formulation of the
asymptotic problem, for which the dimensional bare charges entering
into the velocity correlator (\ref{Fin}) depend on the UV scale
only; see Eq. (\ref{gg}). In a number of papers, e.g.,
\cite{synth,Pumir2,ShS,ShS2,PDF,Falk3,Eyink}, a different version of the
problem was considered, in which the velocity correlator depends also
on the IR scale $L\equiv m^{-1}$. The RG method is applicable also to
this ``exotic'' (from the viewpoints of the theory of critical behavior)
situation. No additional calculation of the fixed points or
anomalous exponents is required; only the regions of stability
of the fixed points FPI--FPV can change.\footnote{This Section is not
included into the journal version.}

As already said in Sec. \ref{sec:Fixed}, solutions of the RG equation
(\ref{RG1}) can be represented in terms of invariant charges $\bar g$,
$\bar u$. The relations (\ref{exo1}) determine implicitly
the invariant charges as functions of the bare parameters; the former
depend on the scales $\Lambda$, $m$ only through their dependence on
the latter.  The solution of the equations (\ref{exo1}) for any given
$g_{0}$, $u_{0}$ would show which of the fixed points is ``chosen
by the RG flow.'' However, the complete solution of this
problem is too difficult to achieve; the main obstacle is that the
renormalization constants in Eq. (\ref{exo1}) must be taken in the
one-loop order of the RG, while we know them only in the one-loop
order of the ordinary perturbation theory; see Eq. (\ref{Z}). In other
words, we know the renormalization constants only up to the terms
of order $O(g)$, and here we need to take into account all the terms
of the form $(g/\eps)^{n}$ etc.

We shall not try to solve the problem completely, we rather give a
simple recipe that allows one to distinguish between the {\it
nontrivial} scaling regimes. The renormalization constants can be
eliminated from Eq. (\ref{exo1}) using the relation (\ref{svaz});
this gives
\begin{equation}
\chi(k)\equiv\frac{\bar g}{\bar u^{3}} =
\frac{g_{0}}{u^{3}_{0}}\, k^{\eta-\zeta},
\label{exo2}
\end{equation}
with $\zeta\equiv \eps-\eta$, see Eq. (\ref{RC1}).
Of course, the identity (\ref{exo2}) contains less information than
the full system (\ref{exo1}); in particular, we cannot judge from Eq.
(\ref{exo2}) whether or not the invariant charges in the IR regime
are attracted by one of the nontrivial fixed points FPII, FPIV or
FPV. Fortunately, if we know (or assume) that this is the case,
we can definitely say which point is involved. Indeed, for FPII
we have $\bar u\to\infty$, $\bar g/\bar u^{2}=\const$ (see Sec.
\ref{sec:Fixed}) and therefore $\chi(k) \to 0$ at this fixed point.
For FPIV ($\bar u\to0$, $\bar g/\bar u=\const$) we have
$\chi(k) \to \infty$ and for FPV the function
$\chi(k)$ remains $O(1)$. We also note that for this case
$\chi(k)$ is independent of $k$ owing to the equality $\zeta=\eta$.
Therefore, taking the limit $k\to0$ in (\ref{exo2}) leads to
the exact relation $g_{0}/u^{3}_{0}=g_{*}/u^{3}_{*}$ for FPV,
which was used in Sec. \ref{sec:Fixed} when deriving Eq. (\ref{roots}).

For the standard regime (\ref{gg}) it follows from Eq. (\ref{exo2}) that
$\chi(k) \propto (k/\Lambda)^{\eta-\zeta}$. Then the above considerations
show that in the IR asymptotic region, $k<<\Lambda $, the invariant
charges approach the point FPII for $\zeta-\eta<0$, FPIV for
$\zeta-\eta>0$, and FPV for $\zeta=\eta$, in agreement with the
analysis given in Sec. \ref{sec:Fixed}.

Now let us turn to the example of the ``exotic'' regime
for which the characteristic
frequency of the velocity field depends only on the IR scale,
$\omega\propto {(Wk^{2})}^{1/3} (k/m)^{4/3-\eta}$
[the dependence on the mean energy dissipation rate, $W$,
is implied], while the equal-time second-order structure function
of the velocity is expressed only through the UV scale,
$S_{2}(r)\propto (Wr)^{2/3} (\Lambda r)^{-8/3+\zeta+\eta}$.
The comparison with Eq. (\ref{Fin}) gives
$u_{0}\nu_0 \propto {W}^{1/3} (k/m)^{4/3-\eta}$
and $g_{0}\nu_0^{2}/u_{0} \propto {W}^{2/3}
\Lambda ^{-8/3+\zeta+\eta}$. From Eq. (\ref{exo2}) we then obtain
\begin{equation}
\chi(k)\propto (\Lambda /k)^{\alpha} \, (k/m)^{\beta}\, ,
\label{exo3}
\end{equation}
with the exponents
\begin{equation}
\alpha\equiv -8/3+\zeta+\eta, \qquad
\beta \equiv -8/3 +2\eta.
\label{exo4}
\end{equation}
It follows from Eqs. (\ref{exo3}), (\ref{exo4}) that the regions
of stability of the fixed points have changed: the point FPII is
approached if $\alpha$, $\beta<0$, FPIV is approached if $\alpha$,
$\beta>0$, and the nonuniversal regime FPV takes place for the
unique choice $\alpha=\beta=0$ that exactly corresponds to the
Kolmogorov velocity correlator ($\zeta=\eta=4/3$).

The most interesting scaling regime arises when the signs of the
exponents $\alpha$ and $\beta$ are opposite, say, $\alpha>0$,
$\beta<0$. Then near the upper bound of the inertial range, $k>>m$,
$\Lambda \sim k$, we have $\chi(k) <<1$, and the critical dimensions
are given by the rapid-change fixed point FPIV; near the lower bound
of the inertial range, $\Lambda  >>k$, $k\sim m$, we have $\chi(k)
<<1$, and the ``frozen'' point FPV works. Therefore, the same
critical regime is described by two different fixed points and,
consequently, two sets of critical dimensions! It is tempting to
relate this fact to the effect observed experimentally in the
boundary layer: the second-order structure function exhibits two
power-law regimes with a pronounced break in the exponent
\cite{break}. Of course, one should not insist too much on this
interpretation.

 \section{Conclusion} \label {sec:Con}

We have applied the RG and OPE methods to the simple model
describing the advection of a passive scalar by the synthetic
velocity field and in the presence of an imposed linear
mean gradient. The statistics of the velocity is Gaussian,
with a given self-similar correlator with
finite correlation time.

We have shown that the model possesses the RG symmetry, and the
corresponding RG equations have several fixed points. As a result,
the correlation functions of the scalar field in the
inertial-convective range exhibit various types of scaling
behavior: diffusive-type regimes for which the advection can be
treated within the ordinary perturbation theory, and three nontrivial
convection-type regimes for which the correlation functions of the
model reveal anomalous scaling behavior. The stability of the
fixed points (and, therefore, the choice of the scaling regime)
depends on the values of the two exponents $\eps$ and $\eta$,
entering into the velocity correlator.

The explicit asymptotic expressions for the structure functions
and other correlation functions in any space dimension are obtained;
the anomalous exponents are calculated to the first order of the
corresponding $\eps$ expansions. For the first nontrivial regime
the anomalous exponents are the same as in the rapid-change version
of the model; for the second they are the same as in the model with
time-independent (frozen) velocity field. In these regimes, the
anomalous exponents are universal in the sense that they depend only
on the exponents entering into the velocity correlator; what is more,
they depend on the only exponent ($\zeta\equiv\eps-\eta$ and $\eps$)
remaining in the corresponding limit. For the last
regime the exponents are nonuniversal: in principle, they depend also
on the values of the coupling constants. It turns out, however,
that they can reveal the nonuniversality only in the second order
of the $\eps$ expansion.

A serious question is that of the validity of the $\eps$ expansion
and the possibility of the extrapolation of the results, obtained
within the $\eps$ expansions, to the finite values $\eps=O(1)$. In
Refs. \cite{Gat} and \cite{Pumir2}, the agreement
between the nonperturbative results and the $\eps$ expansion has been
established on the example of the triple correlation function in the
rapid-change model. In particular, in \cite{Pumir2} the exponent
$\Delta[3,1]$ (we use the notation introduced in Sec.
\ref{sec:Operators}) has been calculated numerically for all
$0\le\zeta\le2$ within the zero-mode approach. It was shown that for
small $\zeta$, this nonperturbative result agrees with the expansion
in $\zeta$, while for $\zeta=2$ it coincides with the exact analytic
result $\Delta[3,1]=-2$ obtained previously in \cite{SSP}.
In the paper \cite{VM}, the one-dimensional version of the
rapid-change model has been studied both numerically
and analytically within the zero-mode approach;
the analytic expressions for the anomalous exponents obtained
within the $\eps$ expansion have also been found to agree with
the nonperturbative numerical results. Finally, in Ref. \cite{VMF}
the analytic $O(\eps)$ result has been confirmed by the numerical
experiment on the example of the fourth-order structure function
in three dimensions.

In this connection, we also note that a number of exact analytic
results appear to be in agreement with the corresponding $\eps$
expansions: the exponent $\Delta[2,2]$, calculated exactly in
\cite{Falk1}, the exponent for the second-order structure function
of a passively advected magnetic field \cite{MHD}, and the
second-order exponent for a scalar advected by the nonsolenoidal
(``compressible'') velocity field \cite{RG1}; the corresponding
expansions in $\zeta$ (to the order $\zeta^{2}$) have been calculated
within the RG approach in \cite{RG,RG1}. These facts support
strongly the applicability of the $\eps$ expansion, at least for
low-order correlation functions.

In the paper \cite{Kraich2}, a closure-type approximation for
the rapid-change model, the so-called linear ansatz, was
used to derive simple explicit expression for the anomalous
exponents for any $0\le\zeta\le2$, $d$, and $n$, the order of the
structure function. For $\zeta=1$, the predictions of the linear
ansatz appear to be consistent with the numerical simulations
\cite{Kraich3,Galanti,VMF,Zeitak}; they are also in agreement
with some exact relations \cite{Kraich4,Lvov,Galanti}. However,
they do not agree with the results
obtained within the zero-mode and RG approaches in the ranges of
small $\zeta$, $2-\zeta$ or $1/d$. In fact, there is no {\it formal}
contradiction between the perturbative results and the linear ansatz:
the violation of the latter in the aforementioned limits can be
related to the fact that they have strongly nonlocal dynamics in the
momentum space; see, e.g., the discussion in Refs. \cite{Kraich4,VMF}.
On the other hand, the {\it numerical} divergence of the
predictions given by the linear ansatz and $\eps$ expansion for
the fourth-order structure function at $\zeta \simeq 1$ is, roughly
speaking, of the same order of magnitude as the difference between
the nonperturbative numerical results and the perturbative
small-$\zeta$ results for the triple correlator, as one can see from
the figures presented in \cite{Gat}. One can think that the series in
$\zeta$, obtained within the RG or zero-mode approaches, give correct
formal expansions of the (unknown) exact exponents, while the linear
ansatz gives a good approximate expression for the same quantities
near $\zeta=1$. We also note that the numerical agreement
between the expansion in $\zeta$ and the exact results is expected to
worsen as $n$ increases, because the actual expansion parameter
is $n\zeta$ rather than $\zeta$ itself; see \cite{RG,RG1}.

Of course, it is not impossible that new dangerous operators arise
for some finite value of the RG expansion parameter. The example is
provided by the frozen regime of the model in question (see Sec.
\ref{sec:summ}). In principle, this effect can lead to a qualitative
changeover in the IR behavior of the correlators as $\eps$ increases,
but in our model this is not the case, at least for the second-order
structure function.

We also note that the first-order expressions
for the anomalous exponents in Eq. (\ref{Dnp}) look alike for the
rapid-change and frozen regimes, but we expect that the analytic
properties of the series are different. The aforementioned exact
results suggest that in the rapid-change models, the series in $\zeta$
have finite radii of convergence. In the language of the field theory,
this is related to the fact that in the rapid-change models, there is
no factorial growth of the number of diagrams in higher orders
of the perturbation theory: a great deal of diagrams vanishes
owing to retardation and the fact that the velocity correlator
contains the delta function in time. This is no longer so for the
regimes in which the correlation time remains finite, and we expect
that the series in $\eps$ for these regimes are only asymptotical,
as in most models of the critical behavior.

Let us conclude with a brief discussion of the passive advection in
the non-Gaussian velocity field governed by the nonlinear stochastic
NS equation. In this case, one has to add the nonlinear
term $(v_j \partial_{j})v_{i}$ to the left-hand side of the equation
(\ref{NS}) and set $\eta=0$ in Eq. (\ref{Fin1}). The RG approach is
also applicable to this model; the analysis of the UV divergences
shows that the basic RG functions are the same as for the model
with $\h=0$. The RG analysis of the latter has been accomplished in
\cite{AVH}. It shows that the model possesses a nontrivial IR stable
fixed point;  its coordinate has been calculated in \cite{AVH} in the
first order of the $\eps$ expansion.\footnote{The results obtained
in \cite{AVH} were also rederived later in Refs. \cite{Yakhot2}.}
The inclusion of a nonzero mean gradient $\h\ne0$ gives rise to
anomalous scaling; the analysis given in Sec. \ref{sec:OPE}
can also be extended to this case. For small $\eps$, the
anomalous exponents are given by the relation (\ref{Q}),
in which one should take $g/u(u+1) = \eps/3a$ at the fixed point,
with the coefficient $a$ from Eq. (\ref{a}) (we use the notation
introduced above; the definition of the parameters $\eps$,
$a$, and $u$ in \cite{AVH} is slightly different). Despite the
non-Gaussianity, the critical dimensions of the powers of the
velocity field are given by the simple linear relation
$\Delta[v_{i_1}\cdots v_{i_n}]=n\Delta_{\bf v}=n(1-\eps/3)$;
see \cite{LOMI,JETP,UFN,turbo,She} (in the notation of the papers
\cite{LOMI,JETP,UFN,turbo}, $\eps$ should be replaced with $2\eps$).
Therefore, all these operators are dangerous for $\eps>3$, and
the summation of their contributions is required. For the
different-time correlators, it has been accomplished
in \cite{LOMI,JETP}; for the structure functions it can be performed
in the one-loop approximation as in Sec. \ref{sec:summ} above and
leads to an analogous conclusion: the behavior of the second-order
structure function does not change for $\eps>3$. For $\eps>4$, the
composite operator of the local energy dissipation rate also becomes
dangerous \cite{Pismak}, possibly along with all of its powers
\cite{She}; some other dangerous operators arise for
$\eps>6$ and further \cite{Kim,further}. The identification of all
the other dangerous operators and summation of their contributions
in the operator product expansions remains an open problem.

\acknowledgments

I have benefited from discussions with L.~Ts.~Adzhemyan,
M.~Hnati\v{c}, J.~Honkonen, M.~Yu.~Nalimov and A.~N.~Vasil'ev.
I am thankful to V.~S.~L'vov, A.~Mazzino and M.~Vergassola for
useful comments on the paper subject.

The work was supported by the Russian Foundation for Fundamental
Research (Grant No. 99-02-16783) and by the Grant Center for
Natural Sciences of the Russian State Committee for Higher Education
(Grant No. 97-0-14.1-30).

\vskip2cm

\areas

\begin{table}
\caption{Canonical dimensions of the fields and parameters in the
model (\protect\ref{action}).}
\label{table1}
\begin{tabular}{ccccccccc}
$F$ & $\theta $ & $\theta '$ & $ {\bf v} $ & $\nu$, $\nu _{0}$ & $m$,
$M$, $\mu$, $\Lambda$ & $g_{0}$ & $u_{0}$ & $g$, $u$, ${\bf h}$ \\
\tableline
$d_{F}^{k}$ & $-1$ & $d+1$ & $-1$ & $-2$ & 1& $\eps+\eta $ & $\eta $  & 0 \\
$d_{F}^{\omega }$ & 0 & 0 & 1 & 1 & 0 & 0 & 0 & 0 \\
$d_{F}$ & $-1$ & $d+1$ & 1 & 0 & 1 & $\eps+\eta $ & $\eta $  & 0 \\
\end{tabular}
\end{table}
\end{document}